\documentclass[12pt]{article}
\usepackage{latexsym,cite}
\input amssym.def
\input amssym.tex

\makeatletter
\renewcommand{\theequation}{\thesection.\arabic{equation}}
\@addtoreset{equation}{section}
\makeatother                      

\font\msym=msbm10
\def\Com{{\mathop{\hbox{\msym \char '103}}}}

\begin{document}
\begin{titlepage}
\title{\vskip -60pt
{\small
\begin{flushright} 
DAMTP 97-27 \\
hep-th/9703191
\end{flushright}}
\vskip 20pt
$N=1$ Superconformal Symmetry in Four-dimensions \\
~{}\\
~{}\\
~{}}
\author{Jeong-Hyuck Park\thanks{E-mail address:\,J.H.Park@damtp.cam.ac.uk}}
\date{}
\maketitle
\vspace{-1.0cm}
\begin{center}
\textit{Department of Applied Mathematics and Theoretical Physics}\\
\textit{University of Cambridge}\\
\textit{Silver Street, Cambridge, CB3 9EW, England}
\end{center}
\vspace{3.5cm}
\begin{abstract}
The $N=1,~d=4$ superconformal group is studied and its 
representations 
are discussed. Under superconformal transformations, 
left invariant 
derivatives and  some class of superfields, including 
\textsl{supercurrents}, are shown to follow these 
representations.  
In other words,  these superfields are
\textsl{quasi-primary} by analogy with two dimensional 
conformal field
theory. Based on these results, we find the general 
forms of the 
two-point  and    the three-point correlation functions 
of the quasi-primary superfields in a group theoretical 
way. In particular, we show that  the two-point   function 
of the supercurrent is  unique up to a  
constant and the general form of the 
three-point function of the supercurrent has two free parameters. 
\end{abstract}
\thispagestyle{empty}
\end{titlepage}
\newpage
\section{Introduction}
Four dimensional conformal field theories, $\mbox{CFT}_{4}$ are of 
special physical interest for the standard relativistic quantum 
field theories. Quantum field  theories at renormalization group 
fixed points are expected to be 
conformally invariant. $N=4$ super Yang-Mills theory in four 
dimensions was shown to be superconformally invariant since its 
$\beta$-function vanishes, and the  
$\beta$-function is proportional to the trace 
anomaly~\cite{sohniuswest,adlercollins,brink1,mandelstam,howestelle1,howestelle2}.  
A few other four dimensional  supersymmetric conformal field theories, 
$\mbox{SCFT}_{4}$ have been known as well~-~certain $N=1$  
supersymmetric theory~\cite{parkeswest}, $N=2~SU(2)$ 
QCD~\cite{argyresplesser,eguchihori} and some 
$N=2$ models~\cite{howestelle3}.  In particular a 
number of studies on $\mbox{SCFT}_{4}$~\cite{hugh,seiberg}  
have been done with regard to the electro-magnetic 
duality~\cite{olivemontonen} where the duality is realized as 
infrared fixed points of ordinary supersymmetric theories. 
Further work has been done  on correlation functions of  
superfields~\cite{molotkov,aneva1,aneva3,conlongwest1,howewest1,howewest3,howewest4,conlongphd}\footnote{I
 am informed by Prof.~P. West that an analysis of a superconformal
Killing equations was also undertaken by B. Conlong~\cite{conlongphd}
and  that this will be discussed further in~\cite{conlongwest2}.} and
operator product 
expansions in 
$\mbox{SCFT}_{4}$~\cite{howewest2,anselmifgj,anselmijohansen}  based on  the
superconformal Ward identities,  where the
infinitesimal superconformal transformation rules for   superfields
play a crucial  role. 
In this paper, we approach the problem in a group theoretical way, 
investigating not just infinitesimal  superconformal symmetry 
generators but  
finite transformations. This method  enables us to  deal
with non-scalar  superfields in a compact way and so turns out to be
useful to prove 
the uniqueness of the correlation functions and get  results in closed
forms. Our method can be found in some works on  
ordinary conformal field theory~\cite{cft}, but has not, to our
knowledge, been applied in superconformal field theory.\newline
The organization of this paper is the following: In section~2, 
starting from the definition of superconformal group, we  study the 
sufficient and necessary conditions for a supercoordinate 
transformation to be  superconformal. We then derive all the 
superconformal group generators from an infinitesimal version of 
these conditions, namely the fundamental elements of superconformal
group.  They are known as supertranslations, superdilations, super 
Lorentz transformations and superinversion. We find $4\times 4$, 
$2\times 2$ matrix representations of  superconformal group and 
show that they give the transformation rules for the left invariant  
derivatives and for a certain class  of superfields, in particular 
the chiral/anti-chiral  superfields and the supercurrents in   
Wess-Zumino model and also in vector superfield theory. In other words, 
these superfields are  \textsl{quasi-primary} by 
analogy with two dimensional conformal field theory. In  section~3, 
based on these results, we study the general forms of correlation
functions 
of superfields, which follow the representations above under 
superconformal transformations and so  give the superconformal 
invariance property to the correlation 
functions. In particular, we show that  the two-point   function 
of the supercurrent is  unique up to a  
constant and the general form of the 
three-point function of the supercurrent has two free parameters.  
The work presented here is an extension of some results in 
ordinary conformal field theory~\cite{hughpetkou,hughjohanna}  
to superconformal field theory.
Throughout the paper we assume the real Minkowski spacetime, and so
the  real $N=1$ superconformal group, $SU(2,2|1)$ rather than the complex
superconformal group, $SL(4|1;\Com)$. The latter case was discussed in
detail in~\cite{howehartwell}. Nevertheless such a distinction is not
relevant to our main result. 

\section{Superconformal Symmetry in Four-dimensions}
\subsection{Superconformal Group}
Following the notations of  Wess \& Bagger~\cite{wessbagger},
 the $N=1$ supersymmetry algebra is
\begin{equation}
\begin{array}{c}
[P_{\mu }~,~P_{\nu }]=[P_{\mu }~,~Q_{\alpha 
}]=[P_{\mu}~
~\overline{Q}_{\dot{
\alpha}}]=\{Q_{\alpha }~,~Q_{\beta
}\}=\{\overline{Q}_{\dot{\alpha}}~,~\overline{Q}_{\dot{\beta}
}\}=0 \\ 
\\ 
\{Q_{\alpha }~,~\overline{Q}_{\dot{\alpha}}\}
=2\sigma_{\alpha\dot{\alpha}}^{\mu}P_{\mu} 
\end{array}
\label{susyalgebra}
\end{equation}
The element of super group  is  given as
\begin{equation}
g(z)=e^{i(-x^{\mu }P_{\mu }+\theta ^{\alpha 
}Q_{\alpha
}+\overline{Q}_{\dot{\alpha}}\bar{\theta}^{\dot{\alpha}})}
\end{equation}
where 
$z^{A}=x^{\mu},\theta^{\alpha},\bar{\theta}^{\dot{\alpha}}$ 
are supercoordinates with the restriction
\begin{equation}
\theta^{\alpha}={(\bar{\theta}^{\dot{\alpha}})}^{\dagger}
\label{thetarestriction}
\end{equation}
$\alpha $ and $\dot{\alpha}$ indices are raised or 
lowered by the
antisymmetric $2\times 2$ matrix $\epsilon $, 
$\epsilon 
_{12}=\epsilon
^{21}=1$, thus $Q^{\alpha }=\epsilon ^{\alpha \beta 
}Q_{\beta }$ etc. The supersymmetric interval, 
$x_{12}$ between $z_{1}$ and $z_{2}$  
is defined by $g(z_{12})=g^{-1}(z_{2})g(z_{1})$ and 
so has the form 
\begin{equation}
\begin{array}{ll}
\multicolumn{2}{l}{x_{12}=x_{1}-x_{2}-
i\theta_{1}\sigma{\bar{\theta}}_{2}
+i\theta_{2}\sigma{\bar{\theta}}_{1}}\\
{}&{}\\
\theta_{12}=\theta_{1}-\theta_{2}~ & ~
\bar{\theta}_{12}=\bar{\theta}_{1}-\bar{\theta}_{2}
\end{array}
\label{susyinterval}
\end{equation}
Left invariant derivatives  
$D_{\alpha},\overline{D}_{\dot{\alpha}}$ are 
\begin{equation}
\begin{array}{ll}
D_{\alpha}=\partial_{\alpha}
+i\bar{\theta}^{\dot{\alpha}}\partial_{\alpha\dot{\alpha}}=
\partial^{-}_{\alpha}~~&~
\overline{D}_{\dot{\alpha}}=-\bar{\partial}_{\dot{\alpha}}
-i\theta^{\alpha}\partial_{\alpha\dot{\alpha}}=-\bar{\partial}^{+}_{\dot{\alpha}}
\end{array}
\end{equation}
where
$\partial_{\alpha\dot{\alpha}}=\sigma^{\mu}_{\alpha\dot{\alpha}} 
\partial_{\mu}$ and the superscripts,~ $\pm$, in the 
partial 
differential operators mean that $x_{\pm}=x\pm 
i\theta\sigma\bar{\theta}$ are taken to be fixed. 
\newline
Infinitesimal supersymmetric interval,~${\sf w}$ is 
defined from 
eq.(\ref{susyinterval}) as $z_{1}$ goes to $z_{2}$
\begin{equation}
{\sf w}=dx+i\theta\sigma d\bar{\theta}
-id\theta\sigma\bar{\theta}
\end{equation}
\newline
Superconformal group is the subgroup of 
supercoordinate
transformations, $g:~z \rightarrow z^{\prime}$ that 
preserve the 
infinitesimal supersymmetric interval length, ${\sf 
w^{2}}$  up to a 
local
scale factor
\begin{equation}
g:~z\rightarrow z^{\prime}~~~\Rightarrow~~~{{\sf 
w}^{2}}~\rightarrow~{{\sf w^{\prime}}
^{2}}=\Omega^{2}(g;z){{\sf w}^{2}} 
\label{scdef}
\end{equation}
where $\Omega(g;z)$ is a local scale factor.\newline
We consider  continuous superconformal 
transformations as   
superconformal transformations  which satisfy
\begin{equation}
\begin{array}{lll}
\mbox{det}\left(\frac{\partial^{-
}\theta^{\prime}}{\partial  
\theta}\right)\neq 0&\hspace{2cm}&
\mbox{det}\left(\frac{\partial^{+}\bar{\theta}^{\prime}}
{\partial  \bar{\theta}}\right)\neq 0
\end{array}
\label{continuous-sc}
\end{equation}
By the 
restriction~$\theta^{\prime\alpha}=
(\bar{\theta}^{\prime\dot{\alpha}})^{\dagger}$ , one of these two 
inequalities implies 
the other.
Later we will see that  this condition is preserved 
under the 
successive  continuous superconformal transformations 
so that they 
form a group, namely continuous superconformal 
group.~-~It will be 
shown that $\left(\frac{\partial^{-}\theta^{\prime}}{\partial  
\theta}\right)$ is a representation of the continuous 
superconformal 
group.\newline
Now we are at the position to state that 
for a  supercoordinate transformation, 
$g:z\rightarrow z^{\prime}$ 
which satisfies eq.(\ref{continuous-sc}), the 
necessary and 
sufficient conditions for $g$ to be a (continuous) 
superconformal 
transformation are\footnote{In 
section 2.9 of the book by
I. Buchbinder \& S. Kuzenko~\cite{buchbinder}, they take these 
conditions as the definition of $N=1$ superconformal transformation
and then show that the infinitesimal  supersymmetric interval
length,~${\sf w}^{2}$ is 
preserved under
such a transformation.}
\begin{equation}
\begin{array}{lll}
x^{\prime}_{+}(x_{+},\theta),&\theta^{\prime}(x_{+},\theta)~~
&\mbox{functions of $x_{+}$ and $\theta$ only}\\
{}&{}&{}\\
x^{\prime}_{-}(x_{-
},\bar{\theta}),&\bar{\theta}^{\prime}(x_{-
},\bar{\theta})~~&\mbox{functions of $x_{-}$ and 
$\bar{\theta}$ 
only}
\end{array}
\label{continuous}
\end{equation}
with the reality condition
\begin{equation}
x^{\prime}_{+}(x_{+},\theta)-x^{\prime}_{-}
(x_{-
},\bar{\theta})=2i\theta^{\prime}(x_{+},\theta)\sigma
\bar{\theta}^{\prime}(x_{-},\bar{\theta})
\label{reality}
\end{equation}
This reality condition should be satisfied by the 
definition of 
$x^{\prime}_{\pm}$.\newline
\textit{proof}\newline
Generally under supercoordinate transformation, 
$g:z\rightarrow 
z^{\prime}$, ${\sf w}^{\mu}$ transforms to
\begin{equation}
{\sf w}^{\mu}\rightarrow {\sf 
w^{\prime}}{}^{\mu}=A^{\mu}_{~\nu}(g;z){\sf 
w}^{\nu}+B^{\mu}{}^{\alpha}(g;z)d\theta_{\alpha}
+\bar{B}^{\mu}_{\dot{\alpha}}(g;z)d\bar{\theta}{}^{\dot{\alpha}}
\end{equation}
where
\begin{equation}
A^{\mu}_{~\nu}(g;z)=\frac{\partial 
x^{\prime\mu}_{+}}{\partial x^{\nu}}-2i\frac{\partial 
\theta^{\prime}}{\partial 
x^{\nu}}\sigma^{\mu}\bar{\theta}^{\prime}=\frac{\partial
 x_{-}^{\prime\mu}}{\partial 
x^{\nu}}+2i\theta^{\prime}\sigma^{\mu}\frac{\partial
\bar{\theta}^{\prime}}{\partial x^{\nu}}
\label{A}
\end{equation}
{}\\
\begin{equation}
\begin{array}{l}
B^{\mu}_{\alpha}(g;z)=\partial^{-}_{\alpha}x_{-}^{\prime\mu}-
2i\theta^{\prime}\sigma^{\mu}\partial^{-
}_{\alpha}\bar{\theta}^{\prime}\\
{}\\
\bar{B}^{\mu}_{\dot{\alpha}}(g;z)=-
\bar{\partial}^{+}_{\dot{\alpha}}x_{+}^{\prime\mu}
+2i\bar{\partial}^{+}_{\dot{\alpha}}{\theta}^{\prime}
\sigma^{\mu}
{\bar{\theta}}^{\prime}=(B^{\mu}_{\alpha}(g;z))^{\dagger}
\end{array}
\end{equation}
{}From the definition of  superconformal group we 
require
\begin{equation}
\begin{array}{c}
A^{\lambda}_{~\mu}\eta_{\lambda\rho}A^{\rho}_{~\nu}\propto
\eta_{\mu\nu}\\
{}\\
B^{\mu}_{\alpha}B_{\mu\beta}
=\bar{B}^{\mu}_{\dot{\alpha}}\bar{B}_{\mu\dot{\beta}} 
=A^{\mu}_{~\nu}B_{\mu\alpha}=A^{\mu}_{~\nu}\bar{B}_{\mu\dot
{\alpha}}
=B^{\mu}_{\alpha}\bar{B}_{\mu\dot{\alpha}}=0
\end{array}
\label{equiv1}
\end{equation}
Simple application of chain rule leads to
\begin{equation}
\frac{\partial x_{+}^{\nu}}{\partial
x_{+}^{\prime}{}^{\lambda}}A^{\mu}_{~\nu}+
\frac{\partial\theta^{\alpha}}{\partial 
x_{+}^{\prime}{}^{\lambda}}B^{\mu}_{\alpha}+
\frac{\partial\bar{\theta}^{\dot{\alpha}}}{\partial 
x_{+}^{\prime}{}^{\lambda}}\bar{B}^{\mu}_{\dot{\alpha}}
=\delta^{\mu}_{~\lambda}
\end{equation}
This makes eq.(\ref{equiv1}) simple.  With matrix 
notation   
\begin{equation}
\begin{array}{c}
A^{t}(g;z)\eta A(g;z)=\Omega^{2}(g;z)\eta \\
{} \\
B^{\mu}_{\alpha}=\bar{B}^{\mu}_{\dot{\alpha}}=0 
\end{array}
\label{equiv2}
\end{equation}
{}From  
$\bar{\partial}^{+}_{\dot{\alpha}}\bar{B}^{\mu}_{\dot{\beta}}+ 
\bar{\partial}^{+}_{\dot{\beta}}\bar{B}^{\mu}_{\dot{\alpha}}=0$ we get
\begin{equation}
\bar{\partial}^{+}_{\dot{\alpha}}\theta^{\prime}{}^{\gamma}
\bar{\partial}^{+}_{\dot{\beta}}\bar{\theta}^{\prime}
{}^
{\dot{\gamma}}+
\bar{\partial}^{+}_{\dot{\beta}}\theta^{\prime}{}^{\gamma}
\bar{\partial}^{+}_{\dot{\alpha}}\bar{\theta}^{\prime}{}^
{\dot{\gamma}}=0
\label{alpha}
\end{equation}
By the assumption,~$ 
\mbox{det}\left(\frac{\partial^{+}\bar{\theta}^{\prime}}
{\partial \bar{\theta}}\right)\neq 0$,  we can 
multiply 
$\left(\frac{\partial^{+}\bar{\theta}^{\prime}}{\partial 
\bar{\theta}}\right)^{-1}$ to eq.(\ref{alpha}) to get
\begin{equation}
\bar{\partial}^{+}_{\dot{\alpha}}\theta^{\prime}{}^{\gamma}
\delta^{\dot{\gamma}}_{\dot{\beta}}+
\bar{\partial}^{+}_{\dot{\beta}}\theta^{\prime}{}^{\gamma}
\delta^{\dot{\gamma}}_{\dot{\alpha}}=0
\end{equation}
Putting $\dot{\alpha}=\dot{\beta}=\dot{\gamma}$ gives
\begin{equation}
\bar{\partial}^{+}_{\dot{\alpha}}\theta^{\prime}{}^{\beta}=0
\end{equation}
This and $\bar{B}^{\mu}_{\dot{\alpha}}=0$ imply that 
$x^{\prime}_{+}$ and $\theta^{\prime}$ are functions 
of $x_{+}$ and 
$\theta$ only. Similarly one can show  that   
$x^{\prime}_{-}$ and 
$\bar{\theta}^{\prime}$ are functions of $x_{-}$ and 
$\bar{\theta}$ 
only. Thus these are necessary conditions for $g$ to 
be 
superconformal. Now we need to show that these 
actually imply 
$A^{t}(g;z)\eta A(g;z)  \propto\eta$.\newline
Acting $\bar{\partial}^{+}_{\dot{\alpha}}\partial^{-}_{\alpha}$ to 
the reality condition~(\ref{reality}) gives 
\begin{equation}
\partial_{\alpha\dot{\alpha}}x^{\prime}_{+}-
2i\partial_{\alpha\dot{\alpha}}\theta^{\prime} 
\sigma\bar{\theta}^{\prime}=
\partial^{-
}_{\alpha}\theta^{\prime}\sigma\bar{\partial}^{+}_{\dot{\alpha}}
\bar{\theta}^{\prime}
\end{equation}
and so
\begin{equation}
A^{\mu}_{~\nu}=-
\textstyle{\frac{1}{2}}\tilde{\sigma}^{\dot{\alpha}\alpha}_{\nu}
\partial^{-}_{\alpha}\theta^{\prime}\sigma^{\mu}\bar{\partial}^{+}_
{\dot{\alpha}}
\bar{\theta}^{\prime}
\end{equation}
\begin{equation}
A^{\lambda}_{~\mu}\eta_{\lambda\rho}A^{\rho}_{~\nu}
=\textstyle{\frac{1}{2}}\tilde{\sigma}_{\mu}^{\dot{\alpha}\alpha}
\tilde{\sigma}_{\nu}^{\dot{\beta}\beta}\partial^{-
}_{\alpha}\theta^{\prime}{}^{\gamma}\partial^{-
}_{\beta}\theta^{\prime}_{\gamma}\bar{\partial}^{+}_{
\dot
{\alpha}}
\bar{\theta}^{\prime}_{\dot{\gamma}}\bar{\partial}^{+
}_
{\dot{\beta}}
\bar{\theta}^{\prime}{}^{\dot{\gamma}}
\end{equation}
where $\tilde{\sigma}^{\mu\dot{\alpha}\alpha}= 
\epsilon^{\alpha\beta}\epsilon^{\dot{\alpha}\dot{\beta}} 
\sigma^{\mu}_{\beta\dot{\beta}}$. However from
$\partial^{-}_{\alpha}\theta^{\prime\gamma}\partial^{-}_{\beta}
\theta^{\prime}_{\gamma}+\partial^{-}_{\beta}\theta^{\prime\gamma}
\partial^{-}_{\alpha}\theta^{\prime}_{\gamma}=0$  we get
\begin{equation}
\begin{array}{ll}
\partial^{-}_{\alpha}\theta^{\prime}{}^{\gamma}\partial^{-}_{\beta}
\theta^{\prime}_{\gamma}=\textstyle{\frac{1}{2}}
\epsilon_{\alpha\beta}\partial^{-\delta}\theta^{\prime\gamma}
\partial^{-}_{\delta}\theta^{\prime}_{\gamma}~~&~
\bar{\partial}^{+}_{\dot{\alpha}}
\bar{\theta}^{\prime}_{\dot{\gamma}}\bar{\partial}^{+}_{\dot{\beta}}
\bar{\theta}^{\prime}{}^{\dot{\gamma}}=\textstyle{\frac{1}{2}}\epsilon_{\dot{\beta}
\dot{\alpha}}\bar{\partial}^{+}_{\dot{\delta}}
\bar{\theta}^{\prime}_{\dot{\gamma}}\bar{\partial}^{+}{}^
{\dot{\delta}}
\bar{\theta}^{\prime}{}^{\dot{\gamma}}
\end{array}
\end{equation}
Using 
$\sigma^{\mu}_{\alpha\dot{\alpha}}\tilde{\sigma}^{\dot{\alpha}
\alpha}_{\nu}=-2\delta^{\mu}_{~\nu}$, we finally have
\begin{equation}  
A^{\lambda}_{~\mu}\eta_{\lambda\rho}A^{\rho}_{~\nu}
=\textstyle{\frac{1}{4}}(\partial^{-
}{}^{\delta}\theta^{\prime}{}^{\gamma}\partial^{-
}_{\delta}\theta^{\prime}_{\gamma}\bar{\partial}^{+}_
{\dot{\delta}}
\bar{\theta}^{\prime}_{\dot{\gamma}}\bar{\partial}^{+
}{}^{\dot{\delta}}
\bar{\theta}^{\prime}{}^{\dot{\gamma}})\eta_{\mu\nu}\propto \eta_{\mu\nu}
\label{Asquare}
\end{equation}
This completes our proof.\newline
There is another type of superconformal 
transformation,  which
satisfies
\begin{equation}
\begin{array}{lll}
\mbox{det}\left(\frac{\partial^{-
}\bar{\theta}^{\prime}}{\partial   
\theta}\right)\neq 0&\hspace{2cm}&
\mbox{det}\left(\frac{\partial^{+}\theta^{\prime}}{\partial 
\bar{\theta}}\right)\neq 0
\end{array}
\label{discrete-sc}
\end{equation}
The necessary and sufficient conditions for such a 
supercoordinate
transformation, $g$ to be a  
superconformal transformation are
\begin{equation}
\begin{array}{lll}
x^{\prime}_{+}(x_{-
},\bar{\theta}),&\theta^{\prime}(x_{-
},\bar{\theta})~~&\mbox{functions of $x_{-}$ and 
$\bar{\theta}$ 
only}\\
{}&{}&{}\\
x^{\prime}_{-
}(x_{+},\theta),&\bar{\theta}^{\prime}(x_{+},\theta)~
~
&\mbox{functions of $x_{+}$ and $\theta$ only}\\
\label{discrete}
\end{array}
\end{equation}
with the reality condition
\begin{equation}
x^{\prime}_{+}(x_{-},\bar{\theta})-x^{\prime}_{-}
(x_{+},\theta)=2i\theta^{\prime}(x_{-
},\bar{\theta})\sigma
\bar{\theta}^{\prime}(x_{+},\theta)
\end{equation}      
We will call this type of superconformal 
transformation
``superinversion-type transformation''.\newline
Infinitesimal, therefore continuous, superconformal 
transformation,  
$g:z\rightarrow z^{\prime}\simeq z+\delta z$  
satisfies the 
infinitesimal  reality condition 
\begin{equation}
h(x,\theta,\bar{\theta})=v(x_{+},\theta)-
2i\lambda(x_{+},\theta)\sigma\bar{\theta}=\bar{v}(x_{
-
},\bar{\theta})+2i\theta \sigma\bar{\lambda}(x_{-
},\bar{\theta})=h^{\dagger}(x,\theta,\bar{\theta})
\label{infi-reality}
\end{equation}
where
\begin{equation}
\begin{array}{lll}
\delta x_{+}=v(x_{+},\theta),&\delta\theta=\lambda 
(x_{+},\theta)~~&\mbox{functions of $x_{+}$ and 
$\theta$ only}\\
{}&{}&{}\\
\delta x_{-}=\bar{v}(x_{-
},\bar{\theta})~~,&\delta\bar{\theta}=\bar{\lambda}
(x_{-},\bar{\theta})&\mbox{functions of $x_{-}$ and 
$\bar{\theta}$ 
only}
\end{array}
\end{equation}
Acting 
$\tilde{\sigma}^{\dot{\alpha}\alpha}_{\mu}\{\partial^
{-
}_{\alpha},\bar{\partial}^{+}_{\dot{\alpha}}\}=-
4i\partial_{\mu}$  
to  eq.(\ref{infi-reality}) leads to
\begin{equation}
\partial_{\mu} h_{\nu}+\partial_{\nu} 
h_{\mu}=(D_{\alpha}\lambda^{\alpha}-
\overline{D}_{\dot{\alpha}}\bar{\lambda}^{\dot{\alpha
}})~\eta_
{\mu\nu}=\textstyle{\frac{1}{2}}\partial\cdot h~\eta_{\mu\nu}
\label{h}
\end{equation}
Acting $\bar{\partial}_{\dot{\alpha}}^{+}$ to this 
gives
\begin{equation}
\partial_{\mu}(\lambda\sigma_{\nu})_{\dot{\alpha}}+ 
\partial_{\nu}(\lambda\sigma_{\mu})_{\dot{\alpha}}= 
\textstyle{\frac{1}{2}}\partial_{\rho}(\lambda\sigma^{\rho})_{\dot{\alpha}}
\eta_{\mu\nu}
\end{equation}
and so
\begin{equation}
\partial_{\mu}v_{\nu}+\partial_{\nu}v_{\mu}=\textstyle{\frac{1}{2}}
(\partial \cdot v) \eta_{\mu\nu} 
\end{equation}
Hence $h,v,\lambda$ should be at most quadratic in $x$, and so 
we can put
\begin{equation}
\begin{array}{l}
v^{\mu}(x_{+},\theta)=v^{\mu}(\theta)+ 
w^{\mu}_{~\nu}(\theta)x^{\nu}_{+}
+d(\theta)x^{\mu}_{+}+(\delta^{\mu}_{\nu}x^{2}_{+}-
2x_{+}^{\mu}x_{+\nu})v^{\nu}(\theta)\\
{}\\
\lambda^{\alpha}(x_{+},\theta)=\lambda^{\alpha}(\theta) 
+\lambda^{\alpha}_{\mu}(\theta)x^{\mu}_{+}  
+\lambda^{\alpha}_{\mu\nu}
(\theta)(\eta^{\mu\nu}x^{2}_{+}-
2x_{+}^{\mu}x_{+}^{\nu})
\end{array}
\label{v}
\end{equation}
After substituting these expressions into the 
infinitesimal reality condition~(\ref{infi-reality}) and then by
imposing the reality condition on the second, first 
and zeroth order 
terms  in $x$ successively we can derive the 
following most general 
solution of the infinitesimal reality condition, all 
the generators 
of continuous superconformal transformations, after a 
bit long tedious calculation.  
\begin{equation}
v^{\mu}(x_{+},\theta)=a^{\mu}+2i\theta\sigma^{\mu}\bar{\xi}
+w^{\mu}_{~\nu}x_{+}^{\nu}+\lambda x^{\mu}_{+}-
2\theta\sigma^{\mu}(x_{+}\cdot\tilde{\sigma})\tilde{\zeta}+
b^{\mu}x_{+}^{2}-2x_{+}^{\mu}b\cdot x_{+}
\label{v-sol}
\end{equation}
\begin{equation}
\lambda^{\alpha}(x_{+},\theta)=\xi^{\alpha}+\textstyle{\frac{1}{2}}(\lambda+i\tau)
\theta^{\alpha}+
\textstyle{\frac{1}{4}}w^{\mu\nu}(\theta\sigma_{\mu}\tilde{\sigma}_{\nu})
^{\alpha}+
2\theta^{2}\zeta^{\alpha}-
i(\tilde{\bar{\zeta}}x_{+}\cdot\tilde{\sigma}
)^{\alpha} +(\theta b\cdot\sigma 
x_{+}\cdot\tilde{\sigma})^{\alpha}
\label{lambda-sol}
\end{equation}
where $\theta ^{2}=\theta^{\alpha}\theta _{\alpha},~\bar{\theta}^{2}=
\bar{\theta}_{\dot{\alpha}}\bar{\theta}^{\dot{\alpha}
}$,~$\tilde{\zeta}=\epsilon\zeta$, etc. 
By integration one can get the following finite 
superconformal 
transformations.
\begin{enumerate}
\item Supertranslations, $\hat{z}\oplus z $
\begin{equation}
\begin{array}{lll}
x^{\prime}=y+x-
i\xi\sigma\bar{\theta}+i\theta\sigma\bar{\xi}
~&~ \theta^{\prime}=\xi+\theta~&~
\bar{\theta}^{\prime}=\bar{\xi}+\bar{\theta}
\end{array}
\end{equation}
where $\hat{z}=y,\xi,\bar{\xi}$. One can check  
$g(\hat{z})g(z)=g(\hat{z}\oplus z)$.
\item Super Lorentz transformations, $L(w:z)$
\begin{equation}
\begin{array}{lll}
x^{\prime}=e^{w}x~&~
\theta^{\prime}={\displaystyle
\theta e^{\frac{1}{4}w^{\mu \nu }\sigma _{\mu 
}\tilde{\sigma}_{\nu}}}~&~
\bar{\theta}^{\prime}={\displaystyle
e^{-\frac{1}{4}w^{\mu \nu }\tilde{\sigma} _{\mu 
}\sigma_{\nu}}\bar{\theta}}
\end{array}
\end{equation}
where $(w)^{\mu}_{~\nu}=w^{\mu}_{~\nu},~w_{\mu\nu}=-
w_{\nu\mu}$.
\item Superdilations, $d(\lambda:z)$
\begin{equation}
\begin{array}{lll}
x^{\prime}=|\lambda| x~&~
\theta^{\prime}=\lambda^{\frac{1}{2}}\theta ~&~
\bar{\theta}^{\prime}=\bar{\lambda}^{\frac{1}{2}}\bar
{\theta}
\end{array}
\end{equation}
where $\lambda^{\frac{1}{2}}$ is an arbitrary complex 
number and 
$\bar{\lambda}=\lambda^{\dagger}$. 
\item Superinversion, $i(z)$\newline
Superinversion,~$i(z)$,  a superinversion-type
transformation as the name indicates, is defined 
as~\cite{buchbinder}
\begin{equation}
\begin{array}{lll}
{\displaystyle 
x^{\prime}_{\pm}=\frac{x_{\mp}}{x_{\mp}^{2}} } ~&~
{\displaystyle
\theta^{\prime}=-i\frac{\tilde{\bar{\theta}}~x_{-}\cdot\tilde{\sigma}}
{x_{-}^{2}} }~&~
{\displaystyle
\bar{\theta}^{\prime}=i\frac{x_{+}\cdot\tilde{\sigma}
\tilde
{\theta}}{
x_{+}^{2}}}
\end{array}
\label{superinversion}
\end{equation}
This definition implies 
$\Omega(i;z)=(x^{2}+\theta^{2}\bar{\theta}^{2})^{-1}$ 
and 
$x^{\prime}=x/(x^{2}+\theta^{2}\bar{\theta}^{2})$.
\end{enumerate}
Special superconformal transformation is defined by 
$i(\hat{z}\oplus i(z))$ 
\begin{equation}
\begin{array}{l}
{\displaystyle
x^{\prime}_{+}=\frac{x_{+}+b_{-
}x^{2}_{+}+2\zeta\sigma 
x_{+}\cdot\tilde{\sigma}\tilde{\theta}}{1+2x_{+}\cdot 
b_{-
}+x_{+}^{2}b_{-}^{2}+4\zeta b_{-}\cdot\sigma 
x_{+}\cdot\tilde{\sigma}\tilde{\theta}-
4\zeta\tilde{\theta}-
8\zeta^{2}\theta^{2}}}\\
{}\\
{\displaystyle
\theta^{\prime}=\frac{\theta-\theta x_{+}\cdot\sigma 
b_{-}\cdot 
\tilde{\sigma}+4\zeta\theta^{2}-
i\tilde{\bar{\zeta}}x_{+}\cdot\tilde{\sigma}-
i\tilde{\bar{\zeta}}b_{-
}\cdot\tilde{\sigma}x_{+}^{2}+4i\theta 
x_{+}\cdot\sigma\bar{\zeta}\zeta}{1+2x_{+}\cdot b_{-
}+x_{+}^{2}b_{-
}^{2}+4\zeta b_{-}\cdot\sigma 
x_{+}\cdot\tilde{\sigma}\tilde{\theta}-
4\zeta\tilde{\theta}-
8\zeta^{2}\theta^{2}}}
\end{array}
\label{specialsuperconformal}
\end{equation}
where $\hat{z}=b,\zeta,\bar{\zeta}$. Infinitesimally 
this 
definition coincides with  eqs.(\ref{v-sol},\ref{lambda-sol}).\newline
It is now clear that continuous superconformal 
transformations and
superinversion-type
transformations are one to one mapped by 
superinversion.
{}From now on we will   call $\{\oplus,L,d,i\}$ the 
fundamental elements of superconformal group - in the 
sense that any superconformal transformation can be
generated by combining them.
\subsection{Superconformal Group Representation}
Under the successive  superconformal transformations, 
$g^{\prime}\circ 
g:z\stackrel{g}{\longrightarrow}z^{\prime}
\stackrel{g^{\prime}}{\longrightarrow}z^{ \prime\prime}$
\newline
using $x_{+}-x_{-}=2i\theta\sigma\bar{\theta}$ and 
chain rule,  
one can show  that if both $g^{\prime}$ and $g$ are 
continuous superconformal 
transformations
\begin{equation}
\begin{array}{ll}
{\displaystyle
\frac{\partial^{-
}\theta^{\prime\prime}{}^{\alpha}}{\partial 
\theta^{\prime}{}^{\beta}}\frac{\partial^{-
}\theta^{\prime}{}^{\beta}}{\partial 
\theta^{\gamma}}=\frac{\partial^{-
}\theta^{\prime\prime}{}^{\alpha}}{\partial 
\theta^{\gamma}}},~~~& 
{\displaystyle
\frac{\partial^{+}\bar{\theta}^{\prime\prime}{}^{\dot
{\alpha}}} 
{\partial 
\bar{\theta}^{\prime}{}^{\dot{\beta}}}\frac{\partial^
{+}
\bar{\theta}^
{\prime}{}^{\dot{\beta}}}{\partial 
\bar{\theta}^{\dot{\gamma}}}=\frac{\partial^{+}
\bar{\theta}^{\prime\prime}{}^{\dot{\alpha}}}{\partial
\bar{\theta}^{\dot{\gamma}}}}
\end{array}
\end{equation}
if $g^{\prime}$, $g$ are continuous superconformal 
and superinversion-type 
transformations respectively
\begin{equation}
\begin{array}{ll}
{\displaystyle
\frac{\partial^{+}\bar{\theta}^{\prime\prime}{}^{\dot{\alpha}}}
{\partial \bar{\theta}^{\prime}{}^{\dot{\beta}}}\frac{\partial^{-}
\bar{\theta}^{\prime}{}^{\dot{\beta}}}{\partial 
\theta^{\alpha}}=\frac{\partial^{-
}\bar{\theta}^{\prime\prime}{}^{\dot{\alpha}}}{\partial 
\theta^{\alpha}}},~~~&
{\displaystyle
\frac{\partial^{-
}\theta^{\prime\prime}{}^{\alpha}}{\partial 
\theta^{\prime}{}^{\beta}}\frac{\partial^{+}\theta^{\prime\beta}}
{\partial  \bar{\theta}^{\dot{\alpha}}}=\frac{\partial^{+}
\theta^
{\prime\prime}{}^{\alpha}}{\partial 
\bar{\theta}^{\dot{\alpha}}}}
\end{array}
\end{equation}
if $g^{\prime}$, $g$ are superinversion-type and 
continuous  superconformal 
transformations respectively
\begin{equation}
\begin{array}{ll}
{\displaystyle
\frac{\partial^{+}\theta^{\prime\prime}{}^{\alpha}}{\partial 
\bar{\theta}^{\prime\dot{\alpha}}}\frac{\partial^{+}
\bar{\theta}^{\prime\dot{\alpha}}}{\partial 
\bar{\theta}^{\dot{\beta}}}=\frac{\partial^{+}\theta^
{\prime\prime}{}^{
\alpha}}{\partial \bar{\theta}^{\dot{\beta}}}},~~~&
{\displaystyle
\frac{\partial^{-
}\bar{\theta}^{\prime\prime}{}^{\dot{\alpha}}}{\partial 
\theta^{\prime}{}^{\alpha}}\frac{\partial^{-
}\theta^{\prime}{}^{\alpha}}{\partial 
\theta^{\beta}}=\frac{\partial^{-
}\bar{\theta}^{\prime\prime}{}^{\dot{\alpha}}}{\partial 
\theta^{\beta}}}
\end{array}
\end{equation}
if both $g^{\prime}$ and $g$  are superinversion-type transformations
\begin{equation}
\begin{array}{ll}
{\displaystyle
\frac{\partial^{+}\theta^{\prime\prime}{}^{\alpha}}{\partial 
\bar{\theta}^{\prime}{}^{\dot{\alpha}}}\frac{\partial^{-}
\bar{\theta}^{\prime}{}^{\dot{\alpha}}}{\partial \theta^{\beta}}=
\frac{\partial^{-}\theta^{\prime\prime}{}^{\alpha}}{\partial 
\theta^{\beta}}},~~~&
{\displaystyle
\frac{\partial^{-}
\bar{\theta}^{\prime\prime}{}^{\dot{\alpha}}}{\partial
\theta^{\prime}{}^{\alpha}}\frac{\partial^{+}\theta^{\prime}{}^
{\alpha}}
{\partial 
\bar{\theta}^{\dot{\beta}}}=\frac{\partial^{+}\bar{\theta}^
{\prime\prime}{}^{\dot{\alpha}}}{\partial 
\bar{\theta}^{\dot{\beta}}}}
\end{array}
\end{equation}
for any superconformal transformation $g^{\prime}$ 
and $g$ 
\begin{equation}
A^{\mu}_{~\nu}(g^{\prime};z^{\prime})A^{\nu}_{~\rho}(g;z)=A^{\mu}_{~\rho}
(g^{\prime}\circ g;z)  
\end{equation}
Thus the followings are representations of 
superconformal 
group.\newline
\begin{itemize}
\item For continuous superconformal transformations
\begin{equation}
\begin{array}{lll}
{\displaystyle 
A(g;z)}~~~~~&{\displaystyle\frac{\partial^{-}
\theta^{\prime}}{\partial\theta}}~~~~~&
{\displaystyle\frac{\partial^{+}\bar{\theta}^{ 
\prime}}{\partial \bar{\theta}}}
\end{array}
\end{equation} 
\item For superinversion-type transformations
\begin{equation}
\begin{array}{lll}
A(g;z)~~~~~&{\displaystyle\frac{\partial^{-
}\bar{\theta}^{\prime}}{\partial 
\theta}}~~~~~&{\displaystyle\frac{\partial^{+}\theta^{\prime}}{ 
\partial \bar{\theta}}}
\end{array}
\end{equation}
\end{itemize}
These are compactified representations of 
superconformal group from 
$6\times 6$ fundamental representation to $4\times 4$ 
or $2\times 
2$. $A^{\mu}_{~\nu}(g;z)$ is the superconformal 
representation for 
the infinitesimal supersymmetric interval  ${\sf 
w}^{\mu}$
\begin{equation}
{\sf w}^{\prime}{}^{\mu}=A^{\mu}_{~\nu}(g;z){\sf 
w}^{\nu}
\end{equation}
and $\frac{\partial^{-}\theta^{\prime}}{\partial 
\theta},\frac{\partial^{+}\bar{\theta}^{\prime}}
{\partial \bar{\theta}}, \frac{\partial^{-}
\bar{\theta}^{\prime}}{\partial\theta},\frac{\partial^{+}\theta^{\prime}}
{\partial \bar{\theta}}$  are  the superconformal 
representations 
for the left invariant derivatives 
$D_{\alpha}=\partial^{-}_{\alpha},\overline{D}_{\dot{\alpha}}=-
\bar{\partial}^{+}_{\dot{\alpha}}$
\begin{eqnarray}
&\left.
\begin{array}{l}
{\displaystyle
D_{\alpha}=\frac{\partial^{-}\theta^{\prime}{}^{\beta}}{\partial 
\theta^{\alpha}}D^{\prime}_{\beta}}\\
{}\\
{\displaystyle
\overline{D}_{\dot{\alpha}}=\frac{\partial^{+}\bar{\theta}^{\prime}{}
^{\dot{\beta}}}
{\partial 
\bar{\theta}^{\dot{\alpha}}}\overline{D}^{\prime}_{\dot{\beta}}}
\end{array}
\right\}&\mbox{for continuous superconformal 
transformation}
\label{continuousD}\\
&{}& \nonumber\\
&\left.
\begin{array}{l}
D_{\alpha}={\displaystyle
-\frac{\partial^{-}\bar{\theta}^{\prime}{}^{\dot{\alpha}}}{\partial 
\theta^{\alpha}}\overline{D}^{\prime}_{\dot{\alpha}}}
\\
{}\\
\overline{D}_{\dot{\alpha}}={\displaystyle
-\frac{\partial^{+}\theta^{\prime}{}^{\alpha}}
{\partial 
\bar{\theta}^{\dot{\alpha}}}D^{\prime}_{\alpha}}
\end{array}
\right\} &\mbox{for superinversion-type   
transformation~}
\label{discreteD}
\end{eqnarray} 
One can also get
\begin{equation}
{\displaystyle
\partial_{\mu}=A^{\nu}_{~\mu}(g;z)
\partial^{\prime}_{\nu}+\frac{\partial 
\theta^{\prime}{}^{\alpha}}{\partial 
x^{\mu}}D^{\prime}_{\alpha}-
\frac{\partial 
\bar{\theta}^{\prime}{}^{\dot{\alpha}}}{\partial 
x^{\mu}}\overline{D}^{\prime}_{\dot{\alpha}}}
\label{partialmu}
\end{equation}
For the fundamental elements of superconformal group  
we have
\begin{enumerate}
\item Supertranslations
\begin{equation}
\begin{array}{lll}
A^{\mu}_{~\nu}=\delta^{\mu}_{~\nu}~~&~~
{\displaystyle
\frac{\partial^{-
}\theta^{\prime}{}^{\alpha}}{\partial 
\theta^{\beta}}=\delta^{\alpha}_{\beta}}~~&~~
{\displaystyle
\frac{\partial^{+}\bar{\theta}^{
\prime}{}^{\dot{\alpha}}}{\partial  
\bar{\theta}^{\dot{\beta}}}=\delta^{\dot{\alpha}}_{\dot{\beta}}}
\end{array}
\end{equation}
\item Super Lorentz transformations
\begin{equation}
\begin{array}{lll}
{\displaystyle
A^{\mu}_{~\nu}=(e^{w})^{\mu}_{~\nu}}~&~
{\displaystyle
\frac{\partial^{-}\theta^{\prime}{}^{\alpha}}{\partial
\theta^{\beta}}=(e^{\frac{1}{4}w^{\mu \nu}\sigma_{\mu}\tilde{\sigma}_{\nu}})_{\beta}^{~\alpha}}
~&~{\displaystyle
\frac{\partial^{+}\bar{\theta}^
{\prime}{}^{\dot{\alpha}}}{\partial  
\bar{\theta}^{\dot{\beta}}}=(e^{-\frac{1}{4}w^{\mu 
\nu 
}\tilde{\sigma} _{\mu 
}\sigma_{\nu}})^{\dot{\alpha}}_{~\dot{\beta}}}
\end{array}
\end{equation}
\item Superdilations
\begin{equation}
\begin{array}{lll}
A^{\mu}_{~\nu}=|\lambda|\delta^{\mu}_{~\nu}
~~&~
{\displaystyle
\frac{\partial^{-
}\theta^{\prime}{}^{\alpha}}{\partial
\theta^{\beta}}=\lambda^{\frac{1}{2}}\delta^{\alpha}_{\beta}}~~&~~
{\displaystyle
\frac{\partial^{+}\bar{\theta}^{\prime}{}^{\dot{\alpha}}}{\partial 
\bar{\theta}^{\dot{\beta}}}=\bar{\lambda}^{\frac{1}{2}}
\delta^{\dot{\alpha}}_{\dot{\beta}}}
\end{array}
\end{equation}
\item Superinversion
\begin{equation}
\begin{array}{cc}
\multicolumn{2}{c}{{\displaystyle
A^{\mu}_{~\nu}=\frac{1}{x^{2}+\theta^{2}\bar{\theta}^
{2}} 
\left( 
\frac{x^{2}-\theta^{2}\bar{\theta}^{2}}{x^{2} 
+\theta^{2}\bar{\theta}^{2}}\delta^{\mu}_{~\nu}-
2\frac{x^{\mu}x_{\nu}}{x^{2} 
+\theta^{2}\bar{\theta}^{2}}+ 
2{\epsilon^{\mu}}_{\nu\lambda\kappa}\frac{\theta\sigma^{\lambda}
\bar{\theta}}{x^{2}}x^{\kappa}\right) }} \\
{}&{}\\
{\displaystyle
\frac{\partial^{-
}\bar{\theta}^{\prime}{}_{\dot{\alpha}}}{\partial 
\theta^{\alpha}}=-
i\frac{x_{+}\cdot\sigma_{\alpha\dot{\alpha}}}{x^{2}_{
-}}}~~~&~~~
{\displaystyle
\frac{\partial^{+}\theta^{\prime}_{\alpha}}{\partial
\bar{\theta}^{\dot{\alpha}}}=i
\frac{x_{-}\cdot\sigma_{\alpha\dot{\alpha}}}{x^{2}_{+}}}
\end{array}
\end{equation}
\end{enumerate}
{}From the property of superdeterminant one can show 
that for continuous superconformal transformation
\begin{equation}
\begin{array}{l}
\mbox{sdet}\left( 
\begin{array}{cc} 
 \frac{\partial x^{\prime}_{+}}{\partial x} & 
\frac{\partial\theta^{
\prime}}{\partial x} \\  
\frac{\partial^{+} x^{\prime}_{+}}{\partial \theta} & 
\frac{\partial^{+}\theta^{ 
\prime}}{\partial \theta} 
\end{array} 
\right) =\mbox{sdet}\left(
\begin{array}{cc}
A&0\\
0&\frac{\partial^{-}\theta^{\prime}}{\partial\theta}
\end{array}
\right)\stackrel{{\mbox{\scriptsize{shortly}}}}{=}
\mbox{sdet}_{c}\\ 
{}\\
\mbox{sdet}\left( 
\begin{array}{cc}
\frac{\partial x^{\prime}_{-}}{\partial x} & 
\frac{\partial\bar{\theta}
^{\prime}}{\partial x} \\ 
\frac{\partial^{-} x^{\prime}_{-}}{\partial 
\bar{\theta}} & 
\frac{\partial^{-}\bar{
\theta}^{\prime}}{\partial \bar{\theta}}
\end{array}
\right)=\mbox{sdet}\left(
\begin{array}{cc}
A&0\\
0&\frac{\partial^{+}\bar{\theta}^{\prime}}{\partial\bar{\theta}}
\end{array}
\right)\stackrel{{\mbox{\scriptsize{shortly}}}}{=}
\overline{\mbox{sdet}}_{c} 
\end{array} 
\end{equation}
for superinversion-type transformation
\begin{equation}
\begin{array}{l}
\mbox{sdet}\left( 
\begin{array}{cc} 
 \frac{\partial x^{\prime}_{-}}{\partial x} & 
\frac{\partial\bar{\theta}^{
\prime}}{\partial x} \\  
\frac{\partial^{+} x^{\prime}_{-}}{\partial \theta} & 
\frac{\partial^{+}\bar{\theta}^{ 
\prime}}{\partial \theta} 
\end{array} 
\right) =\mbox{sdet}\left(
\begin{array}{cc}
A&0\\
0&\frac{\partial^{-
}\bar{\theta}^{\prime}}{\partial\theta}
\end{array}
\right)\stackrel{{\mbox{\scriptsize{shortly}}}}{=}
\mbox{sdet}_{i}\\ 
{}\\
\mbox{sdet}\left( 
\begin{array}{cc}
\frac{\partial x^{\prime}_{+}}{\partial x} & 
\frac{\partial\theta
^{\prime}}{\partial x} \\ 
\frac{\partial^{-} x^{\prime}_{+}}{\partial 
\bar{\theta}} & 
\frac{\partial^{-}\theta^{\prime}}{\partial 
\bar{\theta}}
\end{array}
\right)=\mbox{sdet}\left(
\begin{array}{cc}
A&0\\
0&\frac{\partial^{+}\theta^{\prime}}{\partial\bar{\theta}}
\end{array}
\right)\stackrel{{\mbox{\scriptsize{shortly}}}}{=} 
\overline{\mbox{sdet}}_{i}
\end{array}
\end{equation}
for any superconformal transformation
\begin{equation} 
\mbox{sdet}=(\overline{\mbox{sdet}})^{\dagger}
\end{equation}
The following formulae can be verified for each 
fundamental element 
of  superconformal group   by direct calculation and 
from the fact 
that $A^{\mu}_{~\nu},\frac{\partial^{-}\theta^{\prime}
{}^{\alpha}}{\partial\theta^{\beta}},\frac{\partial^{+}
\bar{\theta}^{\prime}{}^{\dot{\alpha}}}{\partial\bar{\theta}^{\dot{\beta}}}$
 are 
representations of superconformal group, they can be 
generalized to 
any superconformal transformation.
\begin{itemize}
\item For  continuous superconformal transformation
\begin{equation}
\begin{array}{ll}
A^{t}(g;z)\eta A(g;z)
=(\mbox{sdet}_{c}\overline{\mbox{sdet}}_{c})^{\frac{1}
{3}}\eta ~~&~~ \det 
A=(\mbox{sdet}_{c}\overline{\mbox{sdet}}_{c})^{\frac{
2}{3}}
\end{array}
\end{equation}
\newline
\begin{equation}
\begin{array}{ll}
{\displaystyle
\frac{\partial^{-}\theta^{
\prime}{}^{\gamma}}{\partial\theta^{\alpha}}\frac{\partial^{-
}\theta^{
\prime}{}^{\delta}}{\partial\theta^{\beta}} 
\epsilon_{\gamma\delta}=\frac{(
\mbox{sdet}_{c}\overline{\mbox{sdet}}_{c})^{\frac{2}{
3}}} 
{\mbox{sdet}_{c}}
\epsilon_{\alpha\beta}}~~&~~
{\displaystyle
\det\left(\frac{\partial^{-}\theta^{\prime}}{\partial 
\theta}\right)=\frac{(
\mbox{sdet}_{c}\overline{\mbox{sdet}}_{c})^{\frac{2}{
3}}} 
{\mbox{sdet}_{c}}}
\end{array}
\end{equation}
\newline
\begin{equation}
\begin{array}{ll}
{\displaystyle
\frac{\partial^{+}\bar{\theta}^{
\prime}{}^{\dot{\gamma}}}{\partial\bar{\theta}^{\dot{
\alpha}}}
\frac{\partial^{+}\bar{\theta}^{
\prime}{}^{\dot{\delta}}}{\partial\bar{\theta}^{\dot{
\beta}}} 
\epsilon_
{\dot{\gamma}\dot{\delta}}=\frac{(
\mbox{sdet}_{c}\overline{\mbox{sdet}}_{c})^{\frac{2}{
3}}}
{\overline{\mbox{sdet}}_{c}}
\epsilon_{\dot{\alpha}\dot{\beta}}}~~&~~
{\displaystyle
\det\left(\frac{\partial^{+}\bar{\theta}^{\prime}}
{\partial \bar{\theta}}\right)=\frac{(
\mbox{sdet}_{c}\overline{\mbox{sdet}}_{c})^{\frac{2}{
3}}}
{\overline{\mbox{sdet}}_{c}}}
\end{array}
\end{equation}
\newline
\begin{equation}
{\displaystyle
\frac{\partial^{-}\theta^{
\prime}{}^{\beta}}{\partial\theta^{\alpha}} 
\sigma^{\mu}_{\beta\dot{\beta}}
\frac{\partial^{+}\bar{\theta}^{ 
\prime}{}^{\dot{\beta}}}{\partial\bar{\theta}^{\dot{\alpha}}}
=A^{\mu}_{~\nu}(g;z)\sigma^{\nu}_{\alpha\dot{\alpha}}
}
\label{continuousA}
\end{equation}
\item For  superinversion-type transformation
\begin{equation}
\begin{array}{ll}
A^{t}(g;z)\eta A 
(g;z)=(\mbox{sdet}_{i}\overline{\mbox{sdet}}_{i})^{\frac{1}{3}} 
\eta
~~&~~\det A=-
(\mbox{sdet}_{i}\overline{\mbox{sdet}}_{i})^{\frac{2}
{3}}
\end{array}
\end{equation}
\newline
\begin{equation}
\begin{array}{ll}
{\displaystyle
\frac{\partial^{-}\bar{\theta}^{
\prime}{}^{\dot{\alpha}}}{\partial\theta^{\alpha}}\epsilon_
{\dot{\alpha}\dot{\beta}}\frac{\partial^{-
}\bar{\theta}^{
\prime}{}^{\dot{\beta}}}{\partial\theta^{\beta}}=\frac{(
\mbox{sdet}_{i}\overline{\mbox{sdet}}_{i})^{\frac{2}{3}}} 
{\mbox{sdet}_{i}}
\epsilon_{\alpha\beta}}~~&~~
{\displaystyle
\det\left(\frac{\partial^{-
}\bar{\theta}^{\prime}}{\partial 
\theta}\right)=-\frac{(
\mbox{sdet}_{i}\overline{\mbox{sdet}}_{i})^{\frac{2}{
3}}} 
{\mbox{sdet}_{i}}}
\end{array}
\end{equation}
\newline
\begin{equation}
\begin{array}{ll}
{\displaystyle
\frac{\partial^{+}\theta^{
\prime}{}^{\alpha}}{\partial\bar{\theta}^{\dot{\alpha
}}} \epsilon_
{\alpha\beta}\frac{\partial^{+}\theta^{
\prime}{}^{\beta}}{\partial\bar{\theta}^{\dot{\beta}}
}=\frac{(
\mbox{sdet}_{i}\overline{\mbox{sdet}}_{i})^{\frac{2}{
3}}}
{\overline{\mbox{sdet}}_{i}}
\epsilon_{\dot{\alpha}\dot{\beta}}}~~&~~
{\displaystyle
\det\left(\frac{\partial^{+}\theta^{\prime}}{\partial 
\bar{\theta}}\right)=-\frac{(
\mbox{sdet}_{i}\overline{\mbox{sdet}}_{i})^{\frac{2}{3}}}
{\overline{\mbox{sdet}}_{i}}}
\end{array}
\end{equation}
\newline
\begin{equation}
{\displaystyle
\frac{\partial^{-}\bar{\theta}^{
\prime}_{\dot{\beta}}}{\partial\theta^{\alpha}}   
\tilde{\sigma}^{\mu}{}^{\dot{\beta}\beta}\frac{\partial^{+} \theta^{
\prime}_{\beta}}{\partial\bar{\theta}^{\dot{\alpha}}}
=
A^{\mu}_{~\nu}(g;z)\sigma^{\nu}_{\alpha\dot{\alpha}}}
\label{discreteA}
\end{equation}
\end{itemize}
We identify the local scale factor 
\begin{equation}
\Omega(g;z)=(\det A(g;z))^{\frac{1}{4}}=(\mbox{sdet} 
\overline{\mbox{sdet}})^{\frac{1}{6}}
\end{equation}
We define a local Lorentz transformation, ${\cal 
R}(g;z)$ for any 
superconformal transformation, $g$ by
\begin{equation}
\begin{array}{ll}
{\cal R}^{\mu}_{~\nu}(g;z)=\Omega^{-
1}(g;z)A^{\mu}_{~\nu}(g;z)
&\hspace{2cm}{\cal R}^{t}\eta {\cal R}=\eta
\end{array}
\end{equation}
Surely ${\cal R}(g;z)$ is also a representation of  
superconformal 
group. Specially for superinversion we denote
\begin{equation}
{I^{\mu}}_{\nu}(z)={\cal R}^{\mu}_{~\nu}(i;z) = 
\frac{x^{2}-\theta^{2}\bar{\theta}^{2}}{x^{2} 
+\theta^{2}\bar{\theta}^{2}}\delta^{\mu}_{~\nu}-
2\frac{x^{\mu}x_{\nu}}{x^{2} 
+\theta^{2}\bar{\theta}^{2}} 
+2{\epsilon^{\mu}}_{\nu\lambda\kappa}\frac{\theta\sigma^{\lambda}
\bar{\theta}}{x^{2}}x^{\kappa} 
\end{equation}
This expression will be frequently recalled later. 
It is worth to note $I(i(z))=I^{-1}(z)=I(-z)$.

\section{Superconformal Transformation Rules for 
Superfields}
In this section we study  the superconformal 
transformation rules 
for the chiral/anti-chiral superfields and the 
supercurrents in Wess-Zumino 
model and vector superfield theory  respectively.

\subsection{In Wess-Zumino Model}
In Wess-Zumino model
\begin{equation}
-\textstyle{\frac{1}{8}}{\int\int} d^{4}x_{+}d^{2}\theta~ 
\Phi\overline{D}^{2}\overline{\Phi}
=\int d^{4}x~\left( 
\partial_{\mu}\phi\partial^{\mu}\bar{\phi}
+2i\psi\sigma^{\mu}\partial_{\mu}\bar{\psi}\right)
\end{equation}
chiral superfield 
\begin{equation}
\Phi (x_{+},\theta )=\phi (x_{+})+2\theta ^{\alpha }\psi 
_{\alpha
}(x_{+})+\theta ^{2}F(x_{+})
\label{chiral}
\end{equation}
and anti-chiral 
superfield,~$\overline{\Phi}=\Phi^{\dagger}$ 
transform under superconformal 
transformation~$g:z\rightarrow 
z^{\prime}$,  as 
\begin{itemize}
\item for continuous superconformal transformation
\begin{equation}
\begin{array}{l}
\Phi(x_{+},\theta)~\rightarrow~\mbox{sdet}_{c}
^{\frac{1}{3}}\Phi (x^{\prime}_{+},\theta^{\prime})\\
{}\\
\overline{\Phi}(x_{-},\bar{\theta})~
\rightarrow ~\overline{\mbox{sdet}}
_{c}^{\frac{1}{3}}\overline{\Phi} (x^{\prime}_{-
},\bar{\theta}
^{\prime})
\end{array}
\label{general1}
\end{equation}
\item for superinversion-type transformation
\begin{equation}
\begin{array}{l}
\Phi(x_{+},\theta)~\rightarrow~\mbox{sdet}_{i}
^{\frac{1}{3}}\overline{\Phi}  (x^{\prime}_{-
},\bar{\theta}
^{\prime})\\
{}\\
\overline{\Phi}(x_{-},\bar{\theta})~\rightarrow~\overline{\mbox{sdet}}
_{i}^{\frac{1}{3}}\Phi (x^{\prime}_{+},\theta^{\prime})
\end{array}
\label{general2}
\end{equation}
\end{itemize}
These transformations ensure  that every component field  transforms  
properly~\cite{NPB7039,NPB7773} so that  the 
Lagrangian,~$\partial_{\mu}\phi\partial^{\mu}\bar{\phi}
+2i\psi\sigma^{\mu}\partial_{\mu}\bar{\psi}$ transforms 
as a scalar density with weight 1 and so the action does not
change. This gives the superconformal invariance of the correlation 
functions. One can easily check this for each element of
superconformal  group and from the 
fact that superdeterminant is a representation of superconformal
group,  they hold for any superconformal transformation.\newline 
The supercurrent,~$J_{\alpha\dot{\alpha}}$, in Wess-Zumino model 
is~\cite{supercurrent}
\begin{equation}
J_{\alpha\dot{\alpha}}= D_{\alpha}\Phi\overline{D}_{\dot{\alpha}}
\overline{\Phi}+2i\Phi{\stackrel{\leftrightarrow}
{\partial}}_{\alpha\dot{\alpha}}\overline{\Phi}
\label{wess-zuminosupercurrent}
\end{equation}
Equations of motion $
\overline{D}^{2}\overline{\Phi}=0,~~D^{2}\Phi=0$ make 
the supercurrent 
conserved
\begin{equation}
D^{\alpha}J_{\alpha\dot{\alpha}}=0  
\label{superconservation*}
\end{equation}
Correlation functions of the supercurrent also satisfy the 
conservation 
equation. The typical term of a correlation 
function,~$\langle\cdots 
J_{\alpha\dot{\alpha}
}\cdots A\cdots B\cdots\rangle $ is 
\begin{equation}
D_{\alpha}\langle\Phi A\rangle 
\overline{D}_{\dot{\alpha}}\langle\overline{\Phi} 
B\rangle +2i(-
1)^{\# A}\langle\Phi A\rangle\stackrel{\leftrightarrow}
{\partial}_{\alpha\dot{\alpha}}\langle\overline{\Phi} 
B\rangle  \label{typical}
\end{equation}
where $\# A$ is  $+1$ for bosonic $A$ and $-1$ for 
fermionic $A$.
With chirality, $\overline{D}_{\dot{\alpha}}\langle\Phi 
A\rangle =0$ and the 
equation of motion, $D^{2}\langle\Phi A\rangle =0$, one 
can show that  
eq.(\ref{typical}) satisfies the conservation 
equation.\newline
The power one third appearing in 
eqs.(\ref{general1},\ref{general2}) 
enables us to add a superpotential term, $g\Phi^{3}$, to 
the super 
Lagrangian still maintaining  the superconformal 
symmetry to get an 
interacting superconformal field theory, but the 
supercurrent is not 
conserved any more.
\begin{equation}
D^{\alpha}J_{\alpha\dot{\alpha}}\propto 
g\overline{\Phi}^{2}\overline{D}_{\dot{\alpha}}
\overline{\Phi}
\end{equation}
The only non-vanishing two-point correlation function of 
chiral/anti-chiral 
massless free superfields is 
\begin{equation}
\langle\Phi (x_{1+},\theta_{1})\overline{\Phi}(x_{2-},\bar{\theta}_{2})
\rangle  
=\frac{c}{(x_{1+}-x_{2-}-2i\theta_{1}\sigma\bar{\theta}_{2})^{2}} 
=  
\frac{c}{\left(x_{12}+i\theta_{12}\sigma\bar{\theta}_{12
}\right)^{2}}  
\label{2pt}
\end{equation}
where $c=-\frac{1}{2\pi^{2}}$.
Under superconformal  transformation the chiral 
superfield 
transforms as $\Phi(z)\rightarrow\Phi(z^{\prime})$ or 
$\overline{\Phi}(z^{\prime})$ up to a scale
factor, hence eq.(\ref{2pt}) implies that the 
infinitesimal interval 
length,~${\sf w}^{2}$ is invariant up to a scale factor under 
superconformal transformation. Of course, this is consistent with the 
definition of 
 superconformal group~(\ref{scdef}). Superconformal 
symmetry of the two-point correlation function gives
\begin{equation}
\begin{array}{ll}
x^{\prime 2}_{12\pm}=\mbox{sdet}^{\frac{1}{3}}_{c1}\overline{\mbox{sdet}}^{
\frac{1}{3}}_{c2}~x^{2}_{12\pm}~~~&~~\mbox{for 
continuous 
superconformal transformation}\\
{}&{}\\
x^{\prime 
2}_{12\pm}=\mbox{sdet}^{\frac{1}{3}}_{i1}
\overline{\mbox{sdet}}^{\frac{1}{3}}_{i2}~x^{2}_{12\mp}~~
&~~\mbox{for superinversion-type transformation}
\end{array}
\end{equation}
Multiplying $x^{2}_{12+}$ and $x^{2}_{12-}$ gives
\begin{equation}
x^{\prime2}_{12}+\theta^{\prime2}_{12}\bar{\theta}
^{\prime2}_{12} = 
\Omega(g;z_{1})\Omega(g;z_{2})\left(x_{12}^{2}+\theta_{12}^{2}
{\bar{\theta}}_{12}^{2}\right)
\label{Wsquare}
\end{equation}
\newline
Using the superconformal transformation rules for the 
chiral/anti-
chiral superfield and the left invariant derivatives 
one can show that the supercurrent in Wess-Zumino model 
transforms 
under continuous superconformal transformation as
\begin{equation}
\begin{array}{ll}
J_{\alpha\dot{\alpha}}(z)\rightarrow &
s\bar{s}\frac{\partial^{-}\theta^{\prime}{}^{\beta}}{\partial 
\theta^{\alpha}}\frac{\partial^{+}\bar{\theta}^{\prime}{}^{\dot{\beta}}}
{\partial 
\bar{\theta}^{\dot{\alpha}}}J_{\beta\dot{\beta}}(z^{\prime})
+(D_{\alpha}s\overline{D}_{\dot{\alpha}}\bar{s}+2is{\stackrel{
\leftrightarrow}{\partial}}_{\alpha\dot{\alpha}}\bar{s})
\Phi^{\prime}\overline{\Phi}^{\prime}\\
{}&{}\\
{}&+(D_{\alpha}s\frac{\partial^{+}\bar{\theta}^{\prime} 
{}^{\dot{\beta}}}
{\partial \bar{\theta}^{\dot{\alpha}}}-
2is\partial_{\alpha\dot{\alpha}}\bar{\theta}^{\prime} 
{}^{\dot{\beta}})
\bar{s}\Phi^{\prime}\overline{D}^{\prime}_{\dot{\beta}} 
\overline{\Phi}^{\prime}
-(\overline{D}_{\dot{\alpha}}\bar{s}\frac{\partial^{-}
\theta^{\prime}{}^{\beta}}
{\partial 
\theta^{\alpha}}+2i\bar{s}\partial_{\alpha\dot{\alpha}} 
\theta^{\prime}{}^{\beta})s
\overline{\Phi}^{\prime}D^{\prime}_{\beta}\Phi^{\prime}
\end{array}
\label{contipre}
\end{equation}
where $s=\mbox{sdet}_{c}^{\frac{1}{3}},\bar{s}= 
\overline{\mbox{sdet}}_{c}^{\frac{1}{3}}$, and under 
superinversion-type transformation
\begin{equation}
\begin{array}{ll}
J_{\alpha\dot{\alpha}}(z)\rightarrow &
-s\bar{s}\frac{\partial^{-
}\bar{\theta}^{\prime}{}^{\dot{\beta}}}{\partial 
\theta^{\alpha}}\frac{\partial^{+}\theta^{\prime}{}^{\beta}}
{\partial\bar{\theta}^{\dot{\alpha}}}J_{\beta\dot{\beta}}(z^{\prime})
+(D_{\alpha}s\overline{D}_{\dot{\alpha}}\bar{s}+2is
{\stackrel{\leftrightarrow}{\partial}}_{\alpha\dot{\alpha}}\bar{s})
\Phi^{\prime}
\overline{\Phi}^{\prime}\\
{}&{}\\
{}&-
(D_{\alpha}s\frac{\partial^{+}\theta^{\prime}{}^{\beta}}
{\partial \bar{\theta}^{\dot{\alpha}}}-
2is\partial_{\alpha\dot{\alpha}}\theta^{\prime}{}^{\beta
}) \bar{s}
\overline{\Phi}^{\prime}D^{\prime}_{\beta}\overline{\Phi}^{\prime}
+(\overline{D}_{\dot{\alpha}}\bar{s}\frac{\partial^{-
}\bar{\theta}^{\prime}{}^{\dot{\beta}}}{\partial 
\theta^{\alpha}}+2i\bar{s}\partial_{\alpha\dot{\alpha}} 
\bar{\theta}^{\prime}{}^{\dot{\beta}})s\Phi^{\prime}
\overline{D}^{\prime}_{\dot{\beta}}\overline{\Phi}^{\prime}
\end{array}
\label{dispre}
\end{equation}
where $s=\mbox{sdet}_{i}^{\frac{1}{3}},\bar{s}=\overline{\mbox
{sdet}}_{i}^{\frac{1}{3}}$. However one can also check\footnote{In 
fact, 
considering the successive superconformal transformations 
one can show 
that eqs.(\ref{continuous-s},\ref{discrete-s}) hold for 
any 
continuous/superinversion-type transformation.} that for each 
fundamental element of superconformal group except 
superinversion
\begin{equation}
\begin{array}{ll}
D_{\alpha}s\frac{\partial^{+}\bar{\theta}^{\prime\dot{\beta}}}
{\partial \bar{\theta}^{\dot{\alpha}}}-
2is\partial_{\alpha\dot{\alpha}}\bar{\theta}^{\prime\dot{\beta}}=0~~&~~ 
\overline{D}_{\dot{\alpha}}\bar{s}\frac{\partial^{-}\theta^{\prime\beta}}{
\partial\theta^{\alpha}}+2i\bar{s}\partial_{\alpha\dot{\alpha}} 
\theta^{\prime\beta}=0\\
{}\\
\multicolumn{2}{c}{D_{\alpha}s\overline{D}_{\dot{\alpha}
}\bar{s}+2is
{\stackrel{
\leftrightarrow}{\partial}}_{\alpha\dot{\alpha}}\bar{s}=
0}
\end{array}
\label{continuous-s}
\end{equation}
and for superinversion
\begin{equation}
\begin{array}{ll}
D_{\alpha}s\frac{\partial^{+}\theta^{\prime}{}^{\beta}}
{\partial \bar{\theta}^{\dot{\alpha}}}-
2is\partial_{\alpha\dot{\alpha}}\theta^{\prime}{}^{\beta
}=0~&~
\overline{D}_{\dot{\alpha}}\bar{s}\frac{\partial^{-
}\bar{\theta}^{\prime}{}^{\dot{\beta}}}
{\partial\theta^{\alpha}}+2i\bar{s}\partial_{\alpha\dot{
\alpha}}
\bar{\theta}
^{\prime}{}^
{\dot{\beta}}=0\\
{}\\
\multicolumn{2}{c}{D_{\alpha}s\overline{D}_{\dot{\alpha}
}\bar{s}+2is
{\stackrel{
\leftrightarrow}{\partial}}_{\alpha\dot{\alpha}}\bar{s}=
0}
\end{array}
\label{discrete-s}
\end{equation}
This gives the following superconformal transformation 
rule 
for the supercurrent in Wess-Zumino model\footnote{This 
superconformal transformation rule is valid for any 
superconformal 
transformation, since $\pm_{g}\Omega^{3}(g;z)  {{\cal 
R}^{-
1\mu}}_{\nu}(g;z)$ is a representation of superconformal 
group.}
\begin{equation}
J^{\mu}(z)\rightarrow \pm_{g}\Omega^{3}(g;z)  {{\cal 
R}^{-
1\mu}}_{\nu}(g;z)J^{\nu}(z^{\prime})
\label{simple}
\end{equation}
where 
$J^{\mu}=-\textstyle{\frac{1}{2}}\tilde{\sigma}^{\mu\dot{\alpha}\alpha}
J_{\alpha\dot{\alpha}}
, 
J_{\alpha\dot{\alpha}}=\sigma_{\mu\alpha\dot{\alpha}}J^{\mu}$~~~ 
and 
\begin{equation}
\pm_{g}=\left\{\begin{array}{ll}
+1&\mbox{for continuous $g$}\\
-1&\mbox{for superinversion-type $g$}
\end{array}\right.
\end{equation}
\subsection{In Vector Superfield Theory}
In vector superfield theory
\begin{equation}
-\textstyle{\frac{1}{8}}\int d^{4}x~(\int 
d^{2}\theta~W^{\alpha}W_{\alpha}+\int 
d^{2}\bar{\theta}~\bar{W}_{\dot{\alpha}}\bar{W}^{\dot{\alpha}}) 
=\int d^{4}x~ 
(\textstyle{\frac{1}{4}}f^{\mu\nu}f_{\mu\nu}+i\lambda\sigma^{\mu}\partial
_{\mu}\bar{\lambda})
\end{equation}
where $f_{\mu\nu}=\partial_{\mu}v_{\nu}-
\partial_{\nu}v_{\mu}$. 
Fermionic chiral superfield,~$W_{\alpha}$
\begin{equation}
W_{\alpha}(x_{+},\theta)=-i\lambda_{\alpha}(x_{+}) 
+\{D(x_{+})\delta^{\beta}_{\alpha}-\textstyle{\frac{1}{2}}i
(\sigma^{\mu}\tilde{\sigma}^{\nu})_{\alpha}^{~\beta} 
f_{\mu\nu}(x_{+})\}\theta_{\beta}+\theta^{2} 
\partial_{\alpha\dot{\alpha}}\bar{\lambda}^{\dot{\alpha}}(x_{+})
\label{Wvector}
\end{equation}
and anti-chiral superfield,~$\overline{W}_{\dot{
\alpha}}=(W_{\alpha})^{\dagger}$ transform under 
continuous 
superconformal transformation as
\begin{equation}
\begin{array}{l}
{\displaystyle
W_{\alpha}(x_{+},\theta)\rightarrow 
\frac{\mbox{sdet}_{c}^{\frac{2}{3}}}{\overline{\mbox{sdet}}
^{\frac{1}{3}}_{c}}\frac{\partial^{-
}\theta^{\prime}{}^{\beta}}{\partial 
\theta^{\alpha}}W_{\beta}(x^{\prime}_{+},\theta^{\prime}
)}\\
{}\\
{\displaystyle
\overline{W}_{\dot{\alpha}}(x_{-
},\bar{\theta})\rightarrow 
\frac{\overline{\mbox{sdet}}^{
\frac{2}{3}}_{c}}{\mbox{sdet}_{c}^{\frac{1}{3}}}\frac{
\partial^{+}
\bar{\theta}^{\prime} {}^{\dot{\beta}}}
{\partial 
\bar{\theta}^{\dot{\alpha}}}\overline{W}_{\dot{\beta}}(x^{\prime}_{-
},\bar{\theta}^{\prime})}
\end{array}
\label{continuousW}
\end{equation}
and under superinversion-type transformation
\begin{equation}
\begin{array}{l}
{\displaystyle
W_{\alpha}(x_{+},\theta)\rightarrow 
\frac{\mbox{sdet}_{i}^{\frac{2}{3}}}{\overline{\mbox{sdet}}
^{\frac{1}{3}}_{i}}\frac{\partial^{-} 
\bar{\theta}^{\prime}{}^
{\dot{\alpha}}}{\partial 
\theta^{\alpha}}\overline{W}_{\dot{\alpha}}(x^{\prime}_{
-
},\bar{\theta}^{\prime})}\\
{}\\
{\displaystyle
\overline{W}_{\dot{\alpha}}(x_{-
},\bar{\theta})\rightarrow 
\frac{\overline{\mbox{sdet}}
^{\frac{2}{3}}_{i}}{\mbox{sdet}
^{\frac{1}{3}}_{i}}\frac{\partial^{+}\theta^{\prime}{}^{
\alpha}}
{\partial 
\bar{\theta}^{\dot{\alpha}}}W_{\alpha}(x^{\prime}_{+}, 
\theta^{\prime})}
\end{array}
\label{discreteW}
\end{equation}
One can check that these transformations leave
the  action invariant.\newline
The supercurrent $\tilde{J}_{\alpha\dot{\alpha}}$ in 
vector 
superfield theory is 
\begin{equation}
\tilde{J}_{\alpha\dot{\alpha}}=W_{\alpha}\overline{W}_{\dot{\alpha}}
\label{vectorsupercurrent}
\end{equation}
Equation of motion $D^{\alpha}W_{\alpha}=0$ makes the 
supercurrent 
conserved, $D^{\alpha}\tilde{J}_{\alpha\dot{\alpha}}= 0$. 
The supercurrent transforms under superconformal 
transformation as
\begin{equation}
\tilde{J}^{\mu}(z)\rightarrow \pm_{g}\Omega^{3}(g;z) 
{{\cal R}^{-1\mu}}_{\nu}(g;z)
\tilde{J}^{\nu}(z^{\prime})
\label{tildeJsct}
\end{equation}

\section{Superconformal Invariance of Correlation  Function}
{}From the result of the previous section we have superconformal 
transformation rule for Konishi 
current,~$\Phi(x_{+},\theta)\overline{\Phi}(x_{-},\bar{\theta})$
\begin{equation}
\Phi(x_{+},\theta)\overline{\Phi}(x_{-},\bar{\theta})~\rightarrow
\Omega^{2}(g;z)\Phi(x^{\prime}_{+},\theta^{\prime})
\overline{\Phi}(x^{\prime}_{-},\bar{\theta}^{\prime})
\end{equation}
Keeping this and the superconformal transformation rule for the
supercurrents~(\ref{simple},\ref{tildeJsct}) in our mind, in this 
section we study the correlation  
functions of superfields in a group theoretical way. We write 
superfield as $\Psi^{i}(z)$  and  assume that $\Psi^{i}(z)$ 
transforms under superconformal transformation, $g:z\rightarrow 
z^{\prime}$, as\footnote{Here we focus on only real valued scale 
factor and exclude so called R-factor.}
\begin{equation}
\Psi^{i}(z)\rightarrow \Omega^{\eta}(g;z)D^{i}_{~j}({\cal R}^{-
1}(g;z))\Psi^{j}(z^{\prime})
\end{equation}
where  $\eta$ is the scale dimension of the superfield  and 
$D^{i}_{~j}({\cal R}^{-1}(g;z))$ is a representation for the local 
Lorentz transformation, ${\cal R}^{-1}(g;z)$ and so under the 
successive superconformal transformations,~$ 
z\stackrel{g}{\longrightarrow}z^{\prime}\stackrel{ 
g^{\prime}}{\longrightarrow}z^{\prime\prime}$
\begin{equation}
D^{i}_{~j}({\cal R}^{-1}(g;z))D^{j}_{~k}({\cal R}^{-
1}(g^{\prime};z^{\prime}))=D^{i}_{~k}({\cal R}^{-1}(g^{\prime}\circ 
g;z))
\end{equation}
Correlation  function has  superconformal symmetry 
\begin{equation}
\begin{array}{l}
\langle \Psi_{1}^{i_{1}}(z_{1})\Psi_{2}^{i_{2}}(z_{2}) \cdots 
\Psi_{n}^{i_{n}}(z_{n})\rangle \\
{}\\
=\prod_{a=1}^{n}\Omega^{\eta_{a}} (g;z_{a}){D^{i_{a}}_{a}}_{j_{a}} 
({\cal R}^{-1}(g;z_{a}))\langle \Psi_{1}^{j_{1}}(z^{\prime}_{1}) 
\Psi_{2}^{j_{2}} (z^{\prime}_{2})
\cdots \Psi_{n}^{j_{n}}(z^{\prime}_{n})\rangle 
\end{array}
\label{Green}
\end{equation}
where the subscript,~$a$ of $\Psi^{i}_{a}$ denotes the type of the 
superfield, i.e. scalar superfield, supervector field, etc.\newline
\subsection{Two-point Correlation  Function of General Superfields}
Here we will show that if two superfields have different scale 
dimensions then the two-point function of them vanishes, and 
if they  belong to a same type and so the scale dimensions are equal 
then  two-point function is of the following general form
\begin{equation}
\langle \Psi^{i}(z_{1})\Psi^{j}(z_{2})\rangle =\displaystyle{ 
\frac{D^{i}_{~ 
i^{\prime}}(I^{-1}(z_{12})){\cal 
H}^{i^{\prime}j}}{(x_{12}^{2}+\theta_{12}^{2} 
\bar{\theta}_{12}^{2})^{\eta}}}
\label{2-pointG}
\end{equation}
where $z_{12}=-z_{2}\oplus z_{1}$,~$\eta$ is the scale dimension of 
the superfield and ${\cal 
H}^{ij}$ is a constant matrix which satisfies the superconformal 
symmetric condition
\begin{equation}
{\cal H}^{ij}=D^{i}_{~i^{\prime}}(L)D^{j}_{~j^{\prime}}(L){\cal 
H}^{i^{\prime}j^{\prime}}
\label{Hrestriction}
\end{equation}
where $L$ is an arbitrary Lorentz transformation.\newline
\textit{proof}\newline
Without loss of generality, using the supertranslational invariance, 
we can put the two-point  function as
\begin{equation}
\langle \Psi_{1}^{i_{1}}(z_{1})\Psi_{2}^{i_{2}}(z_{2})\rangle =
\frac{{D^{i}_{1}}_{i^{\prime}}(I^{-1}(z_{12}))H^{i^{\prime}j}(i(z_{12}))}
{(x^{2}_{12}+\theta^{2}_{12} 
\bar{\theta}^{2}_{12})^{\frac{\eta_{1}+\eta_{2}}{2}}}
\end{equation}
For  any superconformal transformation,~$g:z\rightarrow z^{\prime}$, 
one may show (see appendix A for our derivation)
\begin{equation} 
{\cal R}^{-1}(g;z_{2}) 
I(z^{\prime}_{12}){\cal R}(g;z_{1})=I(z_{12}) 
\label{claim} 
\end{equation} 
where $z^{\prime}_{12}=-z^{\prime}_{2}\oplus   z^{\prime}_{1}$. 
By virtue of this relation 
and eq.(\ref{Wsquare}), the superconformal invariance of the
correlation function~(\ref{Green}) leads to
\begin{equation}
H^{ij}(i(z^{\prime}_{12}))=\left(\frac{\Omega(g;z_{2})}
{\Omega(g;z_{1})}\right)^{\frac{\eta_{1}-
\eta_{2}}{2}}{D^{i}_{1}}_{i^{\prime}}({\cal 
R}(g;z_{2})){D^{j}_{2}}_{j^{\prime}}({\cal  
R}(g;z_{2}))H^{i^{\prime}j^{\prime}}(i(z_{12}))
\label{Hij}
\end{equation}
Let us consider a superconformal transformation, 
$\tilde{g}:z\rightarrow z^{\prime}$ defined by
\begin{equation}
\tilde{g}(z)=z^{\prime}_{2}\oplus   i(\tilde{z}\oplus   i(-
z_{2}\oplus z))
\label{2pointsct}
\end{equation}
where $\tilde{g}(z_{2})=z^{\prime}_{2}$ and $\tilde{z}$ are 
arbitrary. This definition gives  
$i(z^{\prime}_{12})=\tilde{z}\oplus    
i(z_{12})$ and  from the explicit form of special superconformal 
transformation~(\ref{specialsuperconformal}) one can easily check 
that ${\cal R}^{\mu}_{~\nu}(\tilde{g};z_{2})=\delta^{\mu}_{~\nu},~ 
\Omega(\tilde{g};z_{2})=1$. On the other hand, we have
\begin{equation}
\Omega(\tilde{g};z_{1})=\Omega(i;i(z^{\prime}_{12})) 
\Omega(i;z_{12})=\frac{x^{\prime2}_{12}+\theta^{\prime2}_{12} 
\bar{\theta}^{\prime 2}_{12}}{x^{2}_{12}+\theta^{2}_{12} 
\bar{\theta}^{2}_{12}}
\end{equation}
Now we choose $\tilde{z}= i(z_{o})\oplus -i(z_{12})$ to get  
$z_{12}^{\prime}=z_{o}$, where $z_{o}$ is arbitrary but fixed.  Then 
eq.(\ref{Hij}) becomes
\begin{equation}
H^{ij}(i(z_{12}))=\frac{{\cal H}^{ij}}{(x^{2}_{12}+\theta^{2}_{12} 
\bar{\theta}^{2}_{12})^{\frac{\eta_{1}-\eta_{2}}{2}}}
\label{Hconst}
\end{equation}
where ${\cal H}^{ij}=(x_{o}^{2}+\theta_{o}^{2}
\bar{\theta}_{o}^{2})^{\frac{\eta_{1}-\eta_{2}}{2}} 
H^{ij}(i(z_{o}))$ is  constant. Substituting this expression back  
into eq.(\ref{Hij}) gives
\begin{equation}
\Omega^{\eta_{1}-\eta_{2}}(g;z_{2}){D^{i}_{1}}_{i^{\prime}}({\cal 
R}(g;z_{2})){D^{j}_{2}}_{j^{\prime}}({\cal R}(g;z_{2})){\cal 
H}^{i^{\prime}j^{\prime}}={\cal H}^{ij}
\end{equation}
If $\eta_{1}\neq\eta_{2}$ then  for this equation to be true for any 
superconformal transformation, specially for superdilations,  ${\cal 
H}^{i_{1}i_{2}}$ should vanish. This completes our proof.\newline
Specifically in the following, assuming that two superfields have 
the same scale dimension,~$\eta$, we study the two-point correlation  
functions of scalar superfields and supervector fields. 
\subsubsection{Two-point Correlation Function of Scalar Superfield,~$S(z)$}
Obviously it has trivial representation and so the two-point  
function has the following unique form
\begin{equation}
\langle S(z_{1})S(z_{2})\rangle 
=\frac{c}{(x_{12}^{2}+\theta^{2}_{12} 
\bar{\theta}^{2}_{12})^{\eta}} 
\end{equation}
where $c$ is constant.
One example is Konishi 
current,~$S(z)=\Phi(x_{+},\theta)\bar{\Phi}(x_{-
},\bar{\theta}),~\eta=2$ in the interacting Wess-Zumino model. In 
free model we get $c=\frac{1}{4\pi^{4}}$.
\subsubsection{Two-point Correlation Function of Supervector 
Field~(Supercurrent),~$V^{\mu}(z)$}
We consider supervector fields, $V^{\mu}(z)$,  of which the
representation has the  form
\begin{equation}
D^{\mu}_{~\nu}({\cal R}^{-1}(g;z))=\pm_{g}{{\cal R}^{-
1\mu}}_{\nu}(g;z)
\end{equation}
and so  $V^{\mu}(z)$ transforms under superconformal transformation 
as
\begin{equation}
V^{\mu}(z)\rightarrow \pm_{g}\Omega^{\eta}(g;z){{\cal R}^{-
1\mu}}_{\nu}(g;z)V^{\nu}(z^{\prime})
\end{equation}
Superconformal  symmetric condition~(\ref{Hrestriction}) is
\begin{equation}
{\cal H}^{\mu\nu}=L^{\mu}_{~\rho}L^{\nu}_{~\lambda}{\cal 
H}^{\rho\lambda}
\label{HH}
\end{equation}
Infinitesimally this becomes 
\begin{equation}
\eta^{\mu\rho}{\cal H}^{\nu\lambda}-\eta^{\nu\rho}{\cal   
H}^{\mu\lambda} +\eta^{\mu\lambda}{\cal H}^{\rho\nu}-
\eta^{\nu\lambda}{\cal  H}^{\rho\mu}=0
\end{equation}
Contracting with  $\eta_{\nu\lambda}$ gives
\begin{equation}
{\cal H}^{\mu\rho}+3{\cal H}^{\rho\mu}=\eta^{\mu\rho}{\cal  
H}^{\nu}_{~\nu}={\cal H}^{\rho\mu}+3{\cal H}^{\mu\rho}
\end{equation}
Hence ${\cal H}^{\mu\nu}$ is proportional to $\eta^{\mu\nu}$ and so
 the two-point  function of supervector field,~$V^{\mu}(z)$  with  
scale dimension,~$\eta$ has the following unique form
\begin{equation}
\langle V^{\mu}(z_{1})V^{\nu}(z_{2})\rangle =c\frac{I^{\mu\nu}(-
z_{12})} 
{(x_{12}^{2}+\theta^{2}_{12} \bar{\theta}^{2}_{12})^{\eta}}
\end{equation}
where $c$ is constant. We confirmed this result by calculating the 
two-point  functions of supercurrents in Wess-Zumino model and 
vector superfield theory respectively and checking them to coincide. 
We get $c=\frac{24}{\pi^{4}}$ in Wess-Zumino model and  
$c=\frac{1}{2\pi^{4}}$ in vector superfield theory. The conservation
 equation below is satisfied if and only if 
$\eta=3$.
\begin{equation}
D^{\alpha}(z_{1})\langle 
V_{\alpha\dot{\alpha}}(z_{1})V_{\beta\dot{\beta}} 
(z_{2})\rangle =0
\end{equation}
where~  
$V_{\alpha\dot{\alpha}}=\sigma_{\mu\alpha\dot{\alpha}}V^{\mu}$. 

\subsection{Three-Point Correlation Function of General Superfields}
Here we will show that three-point  function of superfields has 
the following general form
\begin{equation}
\displaystyle{\langle 
\Psi_{1}^{i}(z_{1})\Psi_{2}^{j}(z_{2})\Psi_{3}^{k} 
(z_{3})\rangle 
=\frac{{D^{i}_{1}}_{i^{\prime}}(I^{-
1}(z_{13})){D^{j}_{2}}_{j^{\prime}}(I^{-1}(z_{23}))
{\cal H}^{i^{\prime}j^{\prime}k}(Z_{3})}{(x_{12}^{2} 
+\theta^{2}_{12}\bar{\theta}^{2}_{12})^{\delta_{3}} 
(x_{23}^{2}+\theta^{2}_{23} 
\bar{\theta}^{2}_{23})^{\delta_{1}}(x_{13}^{2}+\theta^{2}_{13} 
\bar{\theta}^{2}_{13})^{\delta_{2}}}}
\label{3point1}
\end{equation}
where $\delta_{1}=\textstyle{\frac{1}{2}}(\eta_{2}+\eta_{3}-
\eta_{1}),~\delta_{2}=\textstyle{\frac{1}{2}}(\eta_{3}+\eta_{1}-
\eta_{2}),~\delta_{3}=\textstyle{\frac{1}{2}}(\eta_{1}+\eta_{2}-\eta_{3})$ and
\begin{equation}
Z_{3}=-i(z_{23})\oplus   i(z_{13})
\end{equation}
Explicitly $Z_{3}=X_{3},~\Theta_{3},~\overline{\Theta}_{3}$ have the 
form
\begin{equation}
\begin{array}{c}
X_{3}={\displaystyle
\frac{x_{13}}{x_{13}^{2}+\theta_{13}^{2}\bar{\theta}_{13}^{2}}-
\frac{x_{23}}{x_{23}^{2}+\theta_{23}^{2}\bar{\theta}_{23}^{2}}+ 
i\frac{\theta_{23}x_{23+}\cdot\sigma\tilde{\sigma}x_{13-
}\cdot\sigma\bar{\theta}_{13}}{x_{23+}^{2}x_{13-}^{2}}-
i\frac{\theta_{13}x_{13+}\cdot\sigma\tilde{\sigma}x_{23-
}\cdot\sigma\bar{\theta}_{23}}{x_{13+}^{2}x_{23-}^{2}}}\\
{}\\
\Theta_{3}={\displaystyle
i\frac{\tilde{\bar{\theta}}_{23}x_{23-
}\cdot\tilde{\sigma}}{x^{2}_{23-}}-
i\frac{\tilde{\bar{\theta}}_{13}x_{13-
}\cdot\tilde{\sigma}}{x^{2}_{13-}}}\\
{}\\
\overline{\Theta}_{3}={\displaystyle
i\frac{x_{13+}\cdot\tilde{\sigma}\tilde{\theta}_{13}}{x^{2}_{13+}}-
i\frac{x_{23+}\cdot\tilde{\sigma}\tilde{\theta}_{23}}{x^{2}_{23+}}}
\end{array}
\end{equation}
${\cal H}^{ijk}(z)$ is of the form
\begin{equation}
{\cal H}^{ijk}(z)={\cal H}^{ijk}_{1}(x)+{\cal 
H}^{ijk\mu}(x)\theta\sigma_{\mu} \bar{\theta}+{\cal 
H}^{ijk}_{2}(x)\frac{\theta^{2} \bar{\theta}^{2}}{x^{2}}
\label{Hexpansion}
\end{equation}
which satisfies
\begin{equation}
{\cal H}^{ijk}_{a}(rLx)
={D^{i}_{1}}_{i^{\prime}}(L){D^{j}_{2}}_{j^{\prime}}(L) 
{D^{k}_{3}}_{k^{\prime}}(L){\cal 
H}^{i^{\prime}j^{\prime}k^{\prime}}_{a}(x)
\label{H12}
\end{equation}
\begin{equation}
{\cal H}^{ijk\mu}(rLx)
=\pm_{{\scriptscriptstyle 
L}}\frac{1}{r}{D^{i}_{1}}_{i^{\prime}}(L){D^{j}_{2}}_{j^{\prime}}(L)
{D^{k}_{3}}_{k^{\prime}}(L)L^{\mu}_{~\nu}{\cal 
H}^{i^{\prime}j^{\prime}k^{\prime}\nu}(x)
\label{Hmu}
\end{equation}
where  $r$, $L$ are  arbitrary real number \& Lorentz transformation and 
$\pm_{{\scriptscriptstyle L}}=\det L/|\det L|$.\newline
\textit{proof}\newline
Without loss of generality we can put 
\begin{equation}
\langle \Psi_{1}^{i}(z_{1})\Psi_{2}^{j}(z_{2}) 
\Psi_{3}^{k}(z_{3})\rangle 
=\frac{{D^{i}_{1}}_{i^{\prime}}(I^{-1}(z_{13}))
{D^{j}_{2}}_{j^{\prime}}(I^{-
1}(z_{23}))H^{i^{\prime}j^{\prime}k}(i(z_{13}),i(
z_{23}))}{(x_{12}^{2} 
+\theta^{2}_{12}\bar{\theta}^{2}_{12})^{\delta_{3}} 
(x_{23}^{2}+\theta^{2}_{23} 
\bar{\theta}^{2}_{23})^{\delta_{1}}(x_{13}^{2}+\theta^{2}_{13} 
\bar{\theta}^{2}_{13})^{\delta_{2}}}
\end{equation}
With eq.(\ref{claim}), superconformal invariance of the  correlation 
function leads to
\begin{equation}
H^{ijk}(i(z^{\prime}_{13}),i(z^{\prime}_{23}))
= {D^{i}_{1}}_{i^{\prime}}({\cal
R}(g;z_{3})){D^{j}_{2}}_{j^{\prime}}  
({\cal R}(g;z_{3})){D^{k}_{3}}_{k^{\prime}}({\cal
R}(g;z_{3}))H^{i^{\prime}j^{\prime}k^{\prime}}(i(z_{13}),i(z_{23}))
\end{equation}
We consider a superconformal 
transformation, $\tilde{g}:z\rightarrow z^{\prime}$ defined by
\begin{equation}
\tilde{g}(z)= z_{3}^{\prime}\oplus   i(\tilde{z}\oplus   i(-
z_{3}\oplus z))
\end{equation}
where $z_{3}^{\prime}=\tilde{g}(z_{3}),~\tilde{z}$ are arbitrary. 
$\tilde{z}\oplus   i(z_{13})=i(z^{\prime}_{13})$ and ${\cal 
R}^{\mu}_{~\nu}(\tilde{g};z_{3})=\delta^{\mu}_{~\nu}$ imply the 
supertranslational invariance of $H^{ijk}(z_{1},z_{2})$
\begin{equation}
H^{ijk}(\tilde{z}\oplus   z_{1},\tilde{z} \oplus   
z_{2})=H^{ijk}(z_{1},z_{2})
\end{equation}
Hence we can put
\begin{equation}
{\cal H}^{ijk}(z_{12})=H^{ijk}(z_{1},z_{2})
\end{equation}
which gives the general form of  three-point correlation  
function~(\ref{3point1}). 
${\cal H}^{ijk}(z)$ should satisfy the superconformal symmetric 
condition
\begin{equation}
{\cal H}^{ijk}(Z^{\prime}_{3})
={D^{i}_{1}}_{i^{\prime}}({\cal 
R}(g;z_{3})){D^{j}_{2}}_{j^{\prime}}({\cal 
R}(g;z_{3})){D^{k}_{3}}_{k^{\prime}}({\cal R}(g;z_{3})){\cal 
H}^{i^{\prime}j^{\prime}k^{\prime}}(Z_{3})
\label{calH}
\end{equation}
where $g:z\rightarrow z^{\prime}$ is an arbitrary  superconformal 
transformation and
$Z^{\prime}_{3}=-i(z^{\prime}_{23})\oplus   
i(z^{\prime}_{13})$.\newline
However (nicely enough) $Z_{3}=X_{3},\Theta_{3}, 
\overline{\Theta}_{3}$ transforms to 
$Z_{3}^{\prime}=X^{\prime}_{3},\Theta_{3}^{\prime}, 
\overline{\Theta}_{3}^{\prime}$  in a simple form.\newline  
For any superconformal transformation
\begin{equation}
X^{\prime\mu}_{3}=\Omega^{-1}(g;z_{3}){\cal 
R}^{\mu}_{~\nu}(g;z_{3})X^{\nu}_{3}
\label{X3}
\end{equation}
For continuous superconformal transformation
\begin{equation}
\begin{array}{ll}
{\displaystyle\Theta^{\prime\alpha}_{3}=\frac{\mbox{sdet}^{1/3}_{c}} 
{\overline{\mbox{sdet}}_{c}^{2/3}}\left.\frac{\partial^{-
}\theta^{\prime\alpha}}{\partial 
\theta^{\beta}}\right|_{z_{3}}\Theta^{\beta}_{3}},~~~&~~~ 
{\displaystyle  
\overline{\Theta}^{\prime\dot{\alpha}}_{3}=\frac{\overline{ 
\mbox{sdet}}^{1/3}_{c}}{\mbox{sdet}_{c}^{2/3}}\left.\frac{ 
\partial^{+} \bar{\theta}^{\prime\dot{\alpha}}}{\partial 
\bar{\theta}^{\dot{\beta}}}\right|_{z_{3}} 
\overline{\Theta}^{\dot{\beta}}_{3}}
\end{array}
\label{Theta1}
\end{equation}
For superinversion-type transformation
\begin{equation}
\begin{array}{ll}
{\displaystyle\Theta^{\prime\alpha}_{3}=-
\frac{\mbox{sdet}^{1/3}_{i}} 
{\overline{\mbox{sdet}}_{i}^{2/3}}\left.\frac{\partial^{+} 
\theta^{\prime\alpha}}{\partial 
\bar{\theta}^{\dot{\alpha}}}\right|_{z_{3}} 
\overline{\Theta}^{\dot{\alpha}}_{3}},~~~&~~~{\displaystyle  
\overline{\Theta}^{\prime\dot{\alpha}}_{3}= -
\frac{\overline{\mbox{sdet}}^{1/3}_{i}}{\mbox{sdet}_{i}^{2/3}} 
\left.\frac{\partial^{-} \bar{\theta}^{\prime\dot{\alpha}}}{ 
\partial \theta^{\alpha}}\right|_{z_{3}} \Theta^{\alpha}_{3}}
\end{array}
\label{Theta2}
\end{equation}
and so   $\Theta_{3}\sigma^{\mu}\overline{\Theta}_{3}$ transforms 
like a  pseudo vector at $z_{3}$ under superconformal transformation
\begin{equation}
\Theta_{3}^{\prime}\sigma^{\mu}\overline{\Theta}_{3}^{\prime}= 
\pm_{g} \Omega^{-1}(g;z_{3}){\cal 
R}^{\mu}_{~\nu}(g;z_{3})\Theta_{3}\sigma^{\mu} \overline{\Theta}_{3}
\label{pseudo}
\end{equation}
To verify these, we only need to check for the fundamental elements 
of superconformal group, since superdeterminant and 
$\frac{\partial^{-}\theta^{\prime}}{\partial 
\theta},\frac{\partial^{+} \bar{\theta}^{\prime}}{\partial 
\bar{\theta}},\frac{\partial^{+} \theta^{\prime}}{\partial 
\bar{\theta}},\frac{\partial^{-} \bar{\theta}^{\prime}}{\partial 
\theta}$ are all representations of superconformal group. 
Supertranslations, superdilations and super Lorentz transformation  
cases are straightforward. For superinversion we consider a  
superconformal transformation,~$f:z\rightarrow z^{\prime}$ defined by
\begin{equation}
f(z)=i(-i(z_{3}))\oplus   i(z_{3}\oplus   i(z))
\end{equation}
$Z^{\prime}_{3}$ and $Z_{3}$ are related by
$Z^{\prime}_{3}=-f(i(z_{23})) \oplus   f(i(z_{23}\oplus Z_{3}))$  and from
eq.(\ref{KI+1}) we get
\begin{equation}
x^{\prime}_{-}\cdot\sigma=(z_{3}\oplus   i(z))^{\mu}_{-} 
\sigma_{\mu}\ x_{+}\cdot\tilde{\sigma}\ x_{3+}\cdot\sigma
\end{equation}
\begin{equation}
i\bar{\theta}^{\prime}_{\dot{\alpha}}=(i^{\alpha}(z_{3}\oplus   
i(z))-i^{\alpha}(z_{3}))x^{\prime}_{-
}\cdot\sigma_{\alpha\dot{\alpha}}
\end{equation}
Now direct calculation  leads to
\begin{equation}
f(z)=i(-i(z_{3}))\oplus   
((x_{3}^{2}+\theta^{2}_{3}\bar{\theta}^{2}_{3})I^{\mu}_{s\nu}(z_{3})
x^{\nu},i\frac{x^{2}_{3+}}{x^{2}_{3-}}(\tilde{\bar{\theta}}x_{3-
}\cdot\tilde{\sigma})^{\alpha},       -i\frac{x^{2}_{3-
}}{x^{2}_{3+}}(x_{3+}\cdot\tilde{\sigma}\tilde{\theta})^{\dot{\alpha}})
\label{fs}
\end{equation}
which verifies eqs.(\ref{X3},\ref{Theta2}) for superinversion. 
\newline
Invariance under superdilations restricts the power series 
expansion of ${\cal H}^{ijk}(z)$ in $\theta,\bar{\theta}$ as 
\begin{equation}
{\cal H}^{ijk}(z)={\cal H}^{ijk}_{1}(x)+{\cal 
H}^{ijk\mu}(x)\theta\sigma_{\mu} \bar{\theta}+{\cal 
H}^{ijk}_{2}(x)\frac{\theta^{2} \bar{\theta}^{2}}{x^{2}}
\end{equation}
Therefore ${\cal H}^{ijk}(z)$ is a function of 
$x,\theta\sigma\bar{\theta}$~(or $x_{\pm})$ and the  superconformal 
invariance of ${\cal H}^{ijk}$~(\ref{calH}) is equivalent to 
\begin{equation}
{\cal H}^{ijk}(rLx,\pm_{{\scriptscriptstyle 
L}}rL\theta\sigma\bar{\theta})={D^{i}_{1}}_{i^{\prime}}(L){D^{j}_{2}
}_{j^{\prime}}(L){D^{k}_{3}}_{k^{\prime}}(L){\cal 
H}^{i^{\prime}j^{\prime}k^{\prime}}(x,\theta\sigma\bar{\theta})
\label{invH}
\end{equation}
where $r$ and $L$ are arbitrary real number and Lorentz 
transformation. This completes our proof.\newline
Our expression for the three-point  function~(\ref{3point1}) is 
asymmetric in its treatment of superfields. However if we define
\begin{equation}
Z_{1}=-i(z_{31})\oplus i(z_{21})~~~~~Z_{2}=-i(z_{12})\oplus 
i(z_{32})
\end{equation}
then using eqs.(\ref{KI},\ref{KI+1},\ref{I3}) one can get
\begin{equation}
X_{3\pm}^{\mu}={\displaystyle
\frac{x_{12}^{2}+\theta^{2}_{12}\bar{\theta}^{2}_{12}}{x_{23}^{2}+
\theta^{2}_{23}\bar{\theta}^{2}_{23}} 
(I(z_{13})I(Z_{1}))^{\mu}_{~\nu} X^{\nu}_{1\pm}}
={\displaystyle
\frac{x_{12}^{2}+\theta^{2}_{12}\bar{\theta}^{2}_{12}}{x_{13}^{2} 
+\theta^{2}_{13}\bar{\theta}^{2}_{13}} 
(I(z_{23})I(Z_{2}))^{\mu}_{~\nu}X^{\nu}_{2\pm}}
\label{X312}
\end{equation}
\begin{equation}
\begin{array}{c}
I^{-1}(z_{31})I(Z_{1})I(z_{21})=I(z_{23})\\
{}\\
I^{-1}(z_{12})I(Z_{2})I(z_{32})=I(z_{31})\\
{}\\
I^{-1}(z_{23})I(Z_{3})I(z_{13})=I(z_{12})
\label{IZ123}
\end{array}
\end{equation}
Now  by virtue of superconformal symmetry of ${\cal H}^{ijk}$  one 
can  recover a democratic way of treating superfields as
\begin{equation}
\begin{array}{ll}
{}&D^{i}_{1i^{\prime}}(I^{-1}(z_{13}))D^{j}_{2j^{\prime}}(I^{-
1}(z_{23}))
{\cal H}^{i^{\prime}j^{\prime}k}(X_{3\pm})\\
{}&{}\\
=&D^{j}_{2j^{\prime}}(I^{-1}(z_{21}))D^{k}_{3k^{\prime}}(I^{-
1}(z_{31}))
{\cal F}^{ij^{\prime}k^{\prime}}(X_{1\pm})\\
{}&{}\\
=&D^{i}_{1i^{\prime}}(I^{-1}(z_{12}))D^{k}_{3k^{\prime}}(I^{-
1}(z_{32}))
{\cal G}^{i^{\prime}jk^{\prime}}(X_{2\pm})
\end{array}
\end{equation}
where
\begin{equation}
\begin{array}{l}
{\cal F}^{ijk}(X_{1\pm})=D^{i}_{1i^{\prime}}(I(Z_{1})
D^{k}_{3k^{\prime}}(I(Z_{1})){\cal H}^{i^{\prime}jk^{\prime}}(X_{1\pm})\\
{}\\
{\cal G}^{ijk}(X_{2\pm})=D^{i}_{1i^{\prime}}
(I(Z_{2})){\cal H}^{i^{\prime}jk}(I(Z_{2})X_{2\mp})
\end{array}
\end{equation}
Specially for bosonic superfields belonging to a same type, there 
are additional conditions on ${\cal H}$ due to the invariance of 
Green function under permutations of superfields. \newline
\begin{itemize}
\item for $1\leftrightarrow 2$ 
\begin{equation}
{\cal H}^{ijk}(x,\theta\sigma\bar{\theta})={\cal 
H}^{jik}(-x,\theta\sigma\bar{\theta})
\label{1and2}
\end{equation}
\item for $2\leftrightarrow 3$ 
\begin{equation}
D^{i}_{~i^{\prime}}(I(z))
{\cal H}^{i^{\prime}jk}(x,\theta\sigma\bar{\theta})={\cal 
H}^{ikj}(x,-I(z)\theta\sigma\bar{\theta})
\label{2and3}
\end{equation}
\end{itemize}
\subsubsection{Three-point Correlation Function of Scalar 
Superfield,~$S(z)$}
${\cal H}(z)$ has the form
\begin{equation}
{\cal H}(z)={\cal H}_{1}(x)+{\cal 
H}^{\mu}(x)\theta\sigma_{\mu}\bar{\theta}+ {\cal 
H}_{2}(x)\frac{\theta^{2}\bar{\theta}^{2}}{x^{2}}
\end{equation}
Superconformal invariance of ${\cal H}(z)$ is
\begin{equation}
\begin{array}{ll}
{\cal H}_{a}(rLx)={\cal H}_{a}(x),~~~&~~~{\cal 
H}^{\mu}(rLx)=\pm_{{\scriptscriptstyle L}}{\displaystyle\frac{1}{r}} 
L^{\mu}_{~\nu}{\cal H}^{\nu}(x)
\end{array}
\end{equation}
where $r,L$ are arbitrary real number and Lorentz transformation. 
This implies that  ${\cal H}_{1}(x)$, 
${\cal H}_{2}(x)$ are constant and  ${\cal H}^{\mu}(x)$ is linear in 
$x^{\mu}/x^{2}$ , but under superinversion it should also change the
sign,  therefore ${\cal H}^{\mu}(x)=0$ and so 
three-point  function of scalar superfield has the following 
general form
\begin{equation}
\langle S(z_{1})S(z_{2})S(z_{3})\rangle =\frac{c_{1}+{\displaystyle
c_{2} \frac{\Theta^{2}_{3} 
\overline{\Theta}^{2}_{3}}{X^{2}_{3}}}}{(x_{12}^{2} 
+\theta^{2}_{12}\bar{\theta}^{2}_{12})^{\eta/2} 
(x_{23}^{2}+\theta^{2}_{23} 
\bar{\theta}^{2}_{23})^{\eta/2}(x_{13}^{2}+\theta^{2}_{13} 
\bar{\theta}^{2}_{13})^{\eta/2}}
\end{equation}
Although this expression looks asymmetric, one can show that 
$\frac{\Theta^{2}_{3}\overline{\Theta}^{2}_{3}}{X_{3}^{2}}$ is a 
symmetric quantity under permutations of 
$z_{1},z_{2},z_{3}$.\newline
\textit{proof}\newline
{}From 
\begin{equation}
x_{12-}\cdot\sigma=x_{23+}\cdot\sigma X_{3+}\cdot\sigma x_{13-
}\cdot\sigma
\end{equation}
we get 
\begin{equation}
\begin{array}{cc}
X_{3+}^{2}={\displaystyle
\frac{x_{21+}^{2}}{x_{23+}^{2}x^{2}_{31+}}}
~~~ & ~~~{\displaystyle
X_{3-}^{2}=\frac{x_{21-}^{2}}{x_{23-}^{2}x^{2}_{31-}}}
\end{array}
\end{equation}
hence
\begin{equation}
\begin{array}{l}
X_{3}^{2}+\Theta^{2}_{3}\overline{\Theta}^{2}_{3}={\displaystyle 
\frac{x_{12}^{2} +\theta^{2}_{12}\bar{\theta}^{2}_{12}}{(x^{2}_{23} 
+\theta^{2}_{23}\bar{\theta}^{2}_{23})(x_{31}^{2} 
+\theta^{2}_{31}\bar{\theta}^{2}_{31})}}\\
{}\\
X_{3}^{2}+2\Theta^{2}_{3}\overline{\Theta}^{2}_{3}={\displaystyle 
\textstyle{\frac{1}{2}} (x_{12}^{2} 
+\theta_{12}^{2}\bar{\theta}_{12}^{2})^{2}(\frac{1}{x_{12+}^{2}x_{23
+}^{2}x_{31+}^{2}}+\frac{1}{x_{12-}^{2}x_{23-}^{2}x_{31-}^{2}})}
\end{array}
\label{Xsquare}
\end{equation}
Combining these two shows
\begin{equation}
\begin{array}{ll}
{\displaystyle
\frac{\Theta^{2}_{3}\overline{\Theta}^{2}_{3}}{X_{3}^{2}}}& 
={\displaystyle\textstyle{\frac{1}{2}}(x_{12}^{2} 
+\theta_{12}^{2}\bar{\theta}_{12}^{2})(x_{23}^{2} 
+\theta_{23}^{2}\bar{\theta}_{23}^{2})(x_{31}^{2} 
+\theta_{31}^{2}\bar{\theta}_{31}^{2})(\frac{1}{x_{12+}^{2}x_{23+}^{2}
x_{31+}^{2}}+\frac{1}{x_{12-}^{2}x_{23-}^{2}x_{31-}^{2}})}-1\\
{}&{}\\
{}&={\displaystyle\frac{\theta^{2}_{12}\bar{\theta}^{2}_{12}}{x_{12}
^{2}}+ \frac{\theta^{2}_{23}\bar{\theta}^{2}_{23}}{x_{23}^{2}}+ 
\frac{\theta^{2}_{31}\bar{\theta}^{2}_{31}}{x_{31}^{2}}-
4\frac{\theta_{12}x_{12}\cdot\sigma\bar{\theta}_{12}\theta_{23}x_{23
}\cdot\sigma\bar{\theta}_{23}}{x_{12}^{2}x_{23}^{2}}}\\
{}&{}\\
{}&~{\displaystyle ~-
4\frac{\theta_{23}x_{23}\cdot\sigma\bar{\theta}_{23}\theta_{31}x_{31
}\cdot\sigma\bar{\theta}_{31}}{x_{23}^{2}x_{31}^{2}}-
4\frac{\theta_{31}x_{31}\cdot\sigma\bar{\theta}_{31}\theta_{12}x_{12
}\cdot\sigma\bar{\theta}_{12}}{x_{31}^{2}x_{12}^{2}}}\\
{}&{}\\
{}&={\displaystyle
\frac{\Theta^{2}_{1}\overline{\Theta}^{2}_{1}}{X_{1}^{2}} 
=\frac{\Theta^{2}_{2}\overline{\Theta}^{2}_{2}}{X_{2}^{2}}}
\end{array}
\end{equation}
This completes our proof.\newline
For Konishi current,~$S(z)=\Phi(x_{+},\theta)\bar{\Phi}(x_{-
},\bar{\theta}),~\eta=2$, in massless free Wess-Zumino  model, by 
direct calculation we confirmed this result with  
$c_{1}=c_{2}=\frac{1}{4\pi^{6}}$.
\subsubsection{Three-point Correlation Function of 
Supercurrent,~$V^{\mu}(z)$}
${\cal H}^{\lambda\mu\nu}(z)$ has the form
\begin{equation}
{\cal H}^{\lambda\mu\nu}(z)={\cal H}^{\lambda\mu\nu}_{1}(x)  +{\cal 
H}^{\lambda\mu\nu\kappa}(x)\theta\sigma_{\kappa}\bar{\theta}+ {\cal 
H}^{\lambda\mu\nu}_{2}(x)\frac{\theta^{2}\bar{\theta}^{2}}{x^{2}}
\end{equation}
Superconformal invariance of ${\cal H}^{\lambda\mu\nu}(z)$ is
\begin{equation}
\begin{array}{l}
{\cal H}^{\lambda\mu\nu}_{a}(rLx)
=\pm_{{\scriptscriptstyle 
L}}L^{\lambda}_{~\lambda^{\prime}}L^{\mu}_{~\mu^{\prime}} 
L^{\nu}_{~\nu^{\prime}}{\cal 
H}^{\lambda^{\prime}\mu^{\prime}\nu^{\prime}}_{a}(x)\\
{}\\
{\cal H}^{\lambda\mu\nu\kappa}(rLx)
={\displaystyle\frac{1}{r}}L^{\lambda}_{~\lambda^{\prime}} 
L^{\mu}_{~\mu^{\prime}} 
L^{\nu}_{~\nu^{\prime}}L^{\kappa}_{~\kappa^{\prime}}{\cal 
H}^{\lambda^{\prime}\mu^{\prime}\nu^{\prime}\kappa^{\prime}}(x)
\end{array}
\label{3superconformalinv}
\end{equation}
and so the most general form of ${\cal H}^{\lambda\mu\nu}(z)$ 
is
\begin{equation}
\begin{array}{ll}
{\cal H}^{\lambda\mu\nu}(z)=&{\displaystyle 
c_{1}\epsilon^{\lambda\mu\nu\kappa}  
\frac{x_{\kappa}}{(x^{2})^{\frac{1}{2}}}}\\
{}&{}\\
{}&+h_{1}{\displaystyle\frac{x^{\lambda}x^{\mu}x^{\nu}\theta 
x\cdot\sigma\bar{\theta}}{(x^{2})^{\frac{5}{2}}}+h_{2} 
\frac{x^{\lambda}x^{\mu}\theta\sigma^{\nu}\bar{\theta}} 
{(x^{2})^{\frac{3}{2}}}+h_{3} 
\frac{x^{\mu}x^{\nu}\theta\sigma^{\lambda}\bar{\theta}} 
{(x^{2})^{\frac{3}{2}}}+h_{4} 
\frac{x^{\nu}x^{\lambda}\theta\sigma^{\mu}\bar{\theta}} 
{(x^{2})^{\frac{3}{2}}}}\\
{}&{}\\
{}&+{\displaystyle (h_{5}x^{\lambda}\eta^{\mu\nu}+h_{6} x^{\mu}
\eta^{\nu\lambda}+h_{7} 
x^{\nu}\eta^{\lambda\mu})\frac{\theta
x\cdot\sigma\bar{\theta}}{(x^{2})^{\frac{3}{2}}}}\\
{}&{}\\
{}&+{\displaystyle
(h_{8}\eta^{\lambda\mu}\eta^{\nu\kappa}+h_{9}
\eta^{\mu\nu}\eta^{\lambda\kappa}+h_{10}
\eta^{\nu\lambda}\eta^{\mu\kappa})
\frac{\theta\sigma_{\kappa}\bar{\theta}}{(x^{2})^{\frac{1}{2}}}}\\
{}&{}\\
{}&+{\displaystyle c_{2}\epsilon^{\lambda\mu\nu\kappa}x_{\kappa} 
\frac{\theta^{2}\bar{\theta}^{2}}{(x^{2})^{\frac{3}{2}}}}
\end{array}
\label{Hlmn}
\end{equation}
which can be obtained in the following way: For each point in 
spacetime,~$x$,   by taking a Lorentz transformation and rescaling 
it
properly, one can transform it to a unit vector. 
By considering the infinitesimal action of the \textsl{little} 
Lorentz
group of it as eq.(\ref{3superconformalinv}),
which leaves the unit vector invariant, 
one can get the general forms of
$H^{\kappa\lambda\mu}_{a},~H^{\kappa\lambda\mu\nu}$ 
at the unit vector. Then transforming the unit vector 
back to the original vector,~$x$,  gives the general 
solutions of eq.(\ref{3superconformalinv}) when $L$ is continuous 
Lorentz
transformation. To get the final form~(\ref{Hlmn}) one only needs to 
impose  the extra condition, invariance under   superinversion, on 
them.\newline
For supervector fields belonging to a same type there are two 
additional restrictions
\begin{equation}
\begin{array}{l}
{\cal H}^{\lambda\mu\nu}(x,\theta\sigma\bar{\theta})={\cal 
H}^{\mu\lambda\nu}(-x,\theta\sigma\bar{\theta})\\
{}\nonumber\\
{I^{\kappa}}_{\lambda}(z){\cal 
H}^{\lambda\mu\nu}(x,\theta\sigma\bar{\theta})=-{\cal 
H}^{\kappa\nu\mu}(x,-I(z)\theta\sigma\bar{\theta})
\end{array}
\label{1223}
\end{equation}
these two restrictions imply
\begin{equation}
\begin{array}{ll}
h_{3}=h_{4}~~~&~~~h_{2}+h_{3}+2h_{8}+2c_{1}=0\\
{}&{}\\
h_{5}=h_{6}~~~&~~~h_{6}+h_{7}+2h_{8}+2c_{1}=0\\
{}&{}\\
h_{9}=h_{10}~~~&~~~h_{9}=h_{8}+2c_{1}
\end{array}
\end{equation}
Thus, there exist six independent parameters, including one for 
overall
constant, in the three-point  function of a  supervector field. 
However, if the three-point function satisfies the conservation 
equation
\begin{equation}
0=\frac{\partial~}{\partial 
x_{1}^{\lambda}}\langle V^{\lambda}(z_{1})V^{\mu}(z_{2}) 
V^{\nu}(z_{3})\rangle 
\label{3conservation}
\end{equation}
then one can show that the scale dimension,~$\eta$  should be $3$ 
and the three-point  function has the following general form with two
free parameters, $c,d$
\begin{equation}
\begin{array}{ll}
\langle V^{\lambda}(z_{1})V^{\mu}(z_{2})V^{\nu}(z_{3})\rangle 
=&\displaystyle{
\frac{{I^{\lambda}}_{\kappa}(-z_{13})}{(x_{13}^{2}+\theta^{2}_{13} 
\bar{\theta}^{2}_{13})^{3}}\frac{{I^{\mu}}_{\rho}
(-z_{23})}{(x_{23}^{2}+\theta^{2}_{23}\bar{\theta}^{2}_{23})^{3}}}\\
{}&{}\\
{}&~~~\times\left(c{\cal J}^{\kappa\rho\nu}(Z_{3})+d{\cal
K}^{\kappa\rho\nu}(Z_{3})
\right)
\end{array}
\label{threesupercurrent}
\end{equation}
where
\begin{equation}
Z_{3}=-i(z_{23})\oplus i(z_{13})
\end{equation}
\begin{equation}
\begin{array}{c}
{\displaystyle{\cal J}^{\mu\nu\rho}(z)= \epsilon^{\mu\nu\rho\kappa} 
\frac{x_{\kappa}}{(x^{2})^{2}}-
4\frac{x_{\kappa}x_{\lambda}\theta\sigma_{\tau}\bar{\theta}} 
{(x^{2})^{3}}(\eta^{\mu\rho}{\cal E}^{\kappa\lambda 
,\tau\nu}+\eta^{\nu\rho}{\cal E}^{\kappa\lambda ,\tau\mu}-
\eta^{\mu\nu}{\cal E}^{\kappa\lambda , \tau\rho})
}\\
{}\\
{\cal E}^{\kappa\lambda 
,\mu\nu}=\textstyle{\frac{1}{2}}(\eta^{\kappa\mu}\eta^{\lambda\nu} 
+\eta^{\kappa\nu}\eta^{\lambda\mu})-
\textstyle{\frac{1}{4}}\eta^{\kappa\lambda}\eta^{\mu\nu}
\end{array}
\end{equation}
\newline
\begin{equation}
\begin{array}{ll}
{\cal K}^{\mu\nu\rho}(z)= 
&\displaystyle{\frac{\theta\sigma_{\kappa}\bar{\theta}}{(x^{2})^{4}}}
\left(
(x^{2})^{2}
(\eta^{\mu\nu}\eta^{\rho\kappa}+\eta^{\kappa\mu}\eta^{\nu\rho}
+\eta^{\rho\mu}\eta^{\nu\kappa})\right.\\
{}&{}\\
{}&
+4x^{2}x^{\rho}(x^{\mu}
\eta^{\nu\kappa}+x^{\nu}\eta^{\kappa\mu}+x^{\kappa}\eta^{\mu\nu})\\
{}&{}\\
{}&\left.
-6x^{2}(x^{\mu}x^{\nu}\eta^{\rho\kappa}
+x^{\mu}x^{\kappa}\eta^{\rho\nu}+x^{\nu}x^{\kappa}
\eta^{\rho\mu})-12x^{\mu}x^{\nu}x^{\rho}x^{\kappa}\right)
\end{array}
\end{equation}
${\cal E}^{\kappa\lambda ,\mu\nu}$ is the four dimensional 
projection 
operator, which transforms any $4\times 4$ matrix to traceless and  
symmetric one. 
Furthermore, eq.(\ref{threesupercurrent})  
satisfies the ``strong'' conservation equation
\begin{equation}
0=D^{\alpha}(z_{1})\langle  V_{\alpha\dot{\alpha}}(z_{1}) 
V_{\beta\dot{\beta}}(z_{2})V_{\gamma\dot{\gamma}}(z_{3})\rangle  
\label{strongconservation}
\end{equation}
although one might guess that eq.(\ref{3conservation}) is a weaker  
condition than eq.(\ref{strongconservation}) due to the commutation  
relation $\{D_{\alpha},\overline{D}_{\dot{\alpha}}\}=-
2i\partial_{\alpha\dot{\alpha}}$.  This result is demonstrated in 
appendix B.


\section{Summary \& Discussion}
Supertranslations, superdilations, super Lorentz 
transformations and superinversion are all the 
fundamental elements of the  $N=1$ four dimensional 
superconformal group, in the sense that they generate 
all the  superconformal transformations. There are 
$4\times 4$ and $2\times 2$ representations of 
superconformal group. Under superconformal 
transformations  the left invariant derivatives and 
some class of superfields~(chiral/anti-chiral 
superfields, supercurrents in Wess-Zumino model and 
vector superfield theory) follow these 
representations  without auxiliary terms appearing, 
and so they are quasi-primary.  Due to the 
superconformal symmetric property, the 
two-point correlation function  of supercurrents   
is unique up to a overall constant 
and the general form of the three-point function of supercurrents 
has two free parameters. 
Even if the supercurrent  is not conserved as in 
the interacting theories, the  two-point  function of 
supercurrents is  unique. Readers may refer our  
result  that  two-point function of superfields with 
different scale 
dimensions vanishes and acting derivatives on 
superfields changes the 
scale dimension, although generally the descendent 
fields, differentiated 
quasi-primary fields,  do not  have transformation 
rules in close
forms.  On the other hand, we expect that  the general form of the 
three-point  correlation functions of supercurrents in the
interacting theories  may have  
at most six  independent parameters, including an 
overall constant
but this number can be reduced by considering the 
general forms of the correlation functions  of
supercurrent,~$J^{\mu}$ and its 
divergence,~$\partial_{\mu}J^{\mu}$.  
It would be of interest to get operator product 
expansions of superfields using our results. It is
straightforward to obtain the infinitesimal 
superconformal
transformation rules for superfields and so  Ward 
identities from our
results using such as $\delta\mbox{sdet} M =\delta 
\mbox{str} M$.
\newline
\begin{center}
\large{\textbf{Acknowledgements}}
\end{center}
I am  very grateful to Hugh Osborn for introducing me to this  
mathematically elegant subject and many helpful discussions, in particular 
making available
his own unpublished calculations on superconformal 
algebra which
motivated my group theoretical approach to the 
subject. I also would like to thank Prof.~Peter West for very useful 
discussions and comments.  This work was partly supported by Cambridge
Overseas Trust.


\appendix
\begin{center}
\Large{\textbf{Appendix}}
\end{center}
\section{\large{Superconformal Invariance of $I(z)$}}
Here we derive eq.(\ref{claim})
\begin{equation} 
{\cal R}^{-1}(g;z_{2}) 
I(z^{\prime}_{12}){\cal R}(g;z_{1})=I(z_{12}) 
\label{claima}
\end{equation} 
where $z_{12}=-z_{2}\oplus z_{1},z^{\prime}_{12}=-z^{\prime}_{2}
\oplus z^{\prime}_{1}$ and  $g :z\rightarrow 
z^{\prime}$ is an arbitrary superconformal transformation.\newline
\textit{proof}\newline
It is straightforward to verify this for  supertranslations, super 
Lorentz transformations and superdilations. Superinversion case is 
not simple as usual. Direct calculation using
\begin{equation}
\sigma^{\mu}\tilde{\sigma}^{\lambda}\sigma^{\nu}=\eta^{\mu\nu}
\sigma^{\lambda}-\eta^{\lambda\nu}\sigma^{\mu}-
\eta^{\lambda\mu}\sigma^{\nu}+i\epsilon^{\mu\lambda\nu\rho}
\sigma_{\rho}
\end{equation}
gives 
\begin{equation} 
\begin{array}{l}
x_{+}\cdot\sigma\ \tilde{\sigma}^{\mu}\ x_{-}
\cdot\sigma=(x^{2}+\theta^{2}\bar{\theta}^{2}){I^{\mu} }_{\nu} 
(z) \sigma^{\nu}\\
{}\\
x_{-}
\cdot\sigma\ \tilde{\sigma}^{\mu}\ x_{+}\cdot\sigma=(x^{2}+
\theta^{2}\bar{\theta}^{2}){I^{-1\mu} }_{\nu}(z) \sigma^{\nu}
\end{array}
\label{KI} 
\end{equation} 
and under superinversion one can show
\begin{equation} 
\begin{array}{l}  
x_{1+}\cdot\sigma\ x_{12-}^{\prime}\cdot\tilde{\sigma}\
x_{2-}\cdot\sigma =  x_{12+}\cdot\sigma \\ 
{}\\ 
x_{2+}\cdot\sigma\  x^{\prime}_{12+}\cdot\tilde{\sigma}\ x_{1-}
\cdot\sigma =x_{12-}\cdot\sigma
\end{array}
\label{KI+1}
\end{equation}
These two relations~(\ref{KI},\ref{KI+1})  imply
\begin{equation} 
I^{-1} (z_{2})I (z^{\prime}_{12})I (z_{1})=I (z_{12}) 
\label{I3}
\end{equation} 
Hence eq.(\ref{claim}) holds for each fundamental element of 
superconformal group and also under the successive superconformal 
transformations 
$z\stackrel{g }{\rightarrow}z^{\prime}\stackrel{g^{\prime} } 
{\rightarrow}z^{\prime\prime}$ this equation is preserved.
\begin{equation}
{\cal R}^{-1}(g^{\prime} \circ g ;z_{2}) 
I (z^{\prime\prime}_{12}){\cal R}(g^{\prime} \circ 
g ;z_{1})=I (z_{12}) 
\end{equation}
Thus, eq.(\ref{claima}) holds for any superconformal transformation.
\section{\large{Three-point Correlation Function of Supercurrent}}
Here we show 
first that the three-point  function of supervector fields 
satisfying the strong~(superficially) conservation 
equation~(\ref{strongconservation}) has the 
form~(\ref{threesupercurrent}) and  that the weak~(superficially) 
conservation equation~(\ref{3conservation}) is actually equivalent  
to the strong one.\newline
Using eq.(\ref{Xsquare}) we write the three-point  function as
\begin{equation}
\langle V^{\mu}(z_{1})V^{\nu}(z_{2})V^{\rho}(z_{3})\rangle 
={\displaystyle 
\frac{I^{\mu}_{~\kappa}(-z_{13})I^{\nu}_{~\lambda}(-z_{23})
{\cal H}^{\kappa\lambda\nu}(Z_{3})}{(x_{13}^{2}+\theta^{2}_{13} 
\bar{\theta}^{2}_{13})^{\eta}(x_{23}^{2}+\theta^{2}_{23} 
\bar{\theta}^{2}_{23})^{\eta}(X_{3}^{2}+ 
\Theta^{2}_{3}\overline{\Theta}^{2}_{3})^{\eta /2}
}}
\end{equation}
With $V_{\alpha\dot{\alpha}}=\sigma_{\mu\alpha\dot{\alpha}} V^{\mu}$ 
and    $D_{\alpha}(z_{1})=D_{\alpha}(z_{13})=-
i\frac{x_{13+}\cdot\sigma_{\alpha\dot{\alpha}}}{x_{13-
}^{2}}\overline{D}^{\dot{\alpha}}(Z_{3})$, the ``strong''
conservation 
equation~(\ref{strongconservation}) is equivalent to
\begin{equation}
\begin{array}{ll}
0=&{\displaystyle 
D^{\alpha}(z_{1})\left(\frac{(x_{13+}\cdot\sigma\tilde{\sigma}_{ 
\mu}x_{13-}\cdot\sigma)_{\alpha\dot{\alpha}}}{(x_{13+}^{2}x_{13-
}^{2})^{\frac{\eta +1}{2}}}\frac{{\cal 
H}^{\mu\nu\lambda}(Z_{3})}{(X_{3}^{2}+\Theta_{3}^{2}\overline{ 
\Theta}_{3}^{2})^{\frac{\eta}{2}}}\right)}\\
{}&{}\\
{}=&{\displaystyle 2i(3-\eta)\frac{(\epsilon x_{13-
}\cdot\tilde{\sigma}\sigma_{\mu}\bar{\theta}_{13})_{\dot{\alpha}}}{(
x_{13}^{2}+\theta^{2}_{13}\bar{\theta}_{13}^{2})^{\eta 
+1}}\frac{{\cal 
H}^{\mu\nu\lambda}(Z_{3})}{(X_{3}^{2}+\Theta_{3}^{2}\overline{ 
\Theta}_{3}^{2})^{\frac{\eta}{2}}}}\\
{}&{}\\
{}&{\displaystyle -i\frac{(x_{13+}^{2})^{2}}{(x_{13+}^{2}x_{13-
}^{2})^{ \frac{\eta +3}{2}}}(\tilde{\sigma}_{\mu}x_{13-}\cdot\sigma 
)^{\dot{\beta}}_{~\dot{\alpha}}\overline{D}_{\dot{\beta}}(Z_{3}) 
\frac{{\cal 
H}^{\mu\nu\lambda}(Z_{3})}{(X_{3}^{2}+\Theta_{3}^{2}\overline{ 
\Theta}_{3}^{2})^{\frac{\eta}{2}}}}
\end{array}
\end{equation}
Multiplying $(x_{13+}^{2})^{\frac{\eta -1}{2}}(x_{13-
}^{2})^{\frac{\eta +1}{2}}x_{13-
}\cdot\tilde{\sigma}^{\dot{\alpha}\alpha}$ gives
\begin{equation}
{\displaystyle 2(3-\eta )\frac{x_{13-
}^{2}}{x_{13+}^{2}}(\epsilon\sigma_{\mu}\bar{\theta}_{13})^{ 
\alpha}\frac{{\cal 
H}^{\mu\nu\lambda}(Z_{3})}{(X_{3}^{2}+\Theta_{3}^{2}\overline{ 
\Theta}_{3}^{2})^{\frac{\eta}{2}}}= 
\tilde{\sigma}_{\mu}^{\dot{\alpha} 
\alpha}\overline{D}_{\dot{\alpha}}(Z_{3})\frac{{\cal 
H}^{\mu\nu\lambda}(Z_{3})}{(X_{3}^{2}+\Theta_{3}^{2}\overline{ 
\Theta}_{3}^{2})^{\frac{\eta}{2}}}}
\end{equation}
Left hand side is a function of $z_{13},~Z_{3}$ and right hand side 
is a function of $Z_{3}$ only. For this identity to hold always, 
both sides should vanish. Hence $\eta =3$ and 
\begin{equation}
{\displaystyle
0=\tilde{\sigma}_{\mu}^{\dot{\alpha} 
\alpha}(\bar{\partial}_{\dot{\alpha}}+i(\theta\sigma^{\rho})_{ 
\dot{\alpha}}\partial_{\rho})\frac{{\cal 
H}^{\mu\nu\lambda}(z)}{(x^{2}+\theta^{2} \bar{\theta}^{2})^{3/2}}}
\label{same}
\end{equation}
Explicitly
\begin{equation}
0=\tilde{\sigma}_{\mu}^{\dot{\alpha} 
\alpha}(\bar{\partial}_{\dot{\alpha}}+i(\theta\sigma^{\rho})_{ 
\dot{\alpha}}\partial_{\rho})
\left(\begin{array}{l}
c_{1}\epsilon^{\mu\nu\lambda\kappa}  {\displaystyle 
\frac{x_{\kappa}}{(x^{2})^{2}}}\\
{}\\
+\left(k_{1}x^{\mu}x^{\nu}x^{\lambda}x^{\kappa}+k_{2}x^{2} 
x^{\mu}x^{\nu} 
\eta^{\lambda\kappa}+k_{3}x^{2}x^{\lambda}(x^{\mu}\eta^{\nu\kappa}+x
^{\nu}\eta^{\mu\kappa})\right.\\
{}\\
~+k_{4}x^{2}x^{\kappa}(x^{\mu}  
\eta^{\nu\lambda}+x^{\nu}\eta^{\mu\lambda})+k_{5}x^{2}x^{\kappa} 
x^{\lambda}\eta^{\mu\nu}+ 
k_{6}(x^{2})^{2}\eta^{\mu\nu}\eta^{\lambda\kappa}\\
{}\\
\left.~+k_{7}(x^{2})^{2}(\eta^{\mu\kappa}\eta^{\nu\lambda} 
+\eta^{\mu\lambda} \eta^{\nu\kappa})\right){\displaystyle 
\frac{\theta\sigma_{\kappa}\bar{\theta}}{(x^{2})^{4}}}\\
{}\\
+(c_{2}-\textstyle{\frac{3}{2}}c_{1})\epsilon^{\mu\nu\lambda\kappa}  
x_{\kappa}{\displaystyle 
\frac{\theta^{2}\bar{\theta}^{2}}{(x^{2})^{3}}}
\end{array}\right)
\label{explicitconsv}
\end{equation}
where 
\begin{equation}
\begin{array}{ccc}
2c_{1}=k_{2}+k_{3}+2k_{7}~~&~
2c_{1}=k_{4}+k_{5}+2k_{7}~~&~
2c_{1}=k_{7}-k_{6}
\end{array}
\label{restriction1}
\end{equation}
Eq.(\ref{explicitconsv}) has two sorts of terms, $\theta$-term and 
$\theta^{2}\bar{\theta}$-term, both of which are required to be 
zero.\newline
\newline
For $\theta$-term.\newline
$\theta$-term is of the 
form~$(\theta\sigma^{\mu}\tilde{\sigma}^{\nu})^{\alpha}K_{\mu\nu}(z)=0$.
 Using 
$\sigma_{\alpha\dot{\alpha}}^{\mu}\tilde{\sigma}^{\dot{\beta}\beta}_
{\mu}=-2\delta^{\beta}_{\alpha}\delta^{\dot{\beta}}_{\dot{\alpha}}$~ 
one can easily show that 
$0=\sigma^{\mu}\tilde{\sigma}^{\nu}K_{\mu\nu}(z)$ is equivalent to
\begin{equation}
0=\mbox{Tr}(\sigma^{\kappa}\tilde{\sigma}^{\lambda} 
\sigma^{\mu}\tilde{\sigma}^{\nu})K_{\mu\nu}(z)
=2(\eta^{\kappa\lambda}\eta^{\mu\nu}+\eta^{\kappa\nu} 
\eta^{\lambda\mu}-\eta^{\kappa\mu}\eta^{\lambda\nu}-
i\epsilon^{\kappa\lambda\mu\nu})K_{\mu\nu}(z)
\end{equation}
Hence we have
\begin{equation}
\begin{array}{ll}
0=&(\eta^{\varsigma\tau}\delta^{\kappa}_{\mu}+\eta^{\tau\kappa} 
\delta^{\varsigma}_{\mu}-\eta^{\varsigma\kappa}\delta^{\tau}_{\nu}-
i{\epsilon^{\varsigma\tau\kappa}}_{\mu})\left\{-
ic_{1}\epsilon^{\mu\nu\lambda\rho}((x^{2})^{2} \eta_{\rho\kappa}-
4x^{2}x_{\rho}x_{\kappa})+k_{1}x_{\kappa}x^{\mu}x^{\nu} 
x^{\lambda}\right.\\
{}&{} \\
{}&~+k_{2}x^{2}x^{\mu}x^{\nu}\delta^{\lambda}_{\kappa}+k_{3}x^{2} 
x^{\lambda}(x^{\mu}\delta^{\nu}_{\kappa}+x^{\nu} 
\delta^{\mu}_{\kappa})+k_{4}x^{2}x_{\kappa}(x^{\mu} 
\eta^{\nu\lambda}+x^{\nu}\eta^{\mu\lambda})+k_{5}x_{\kappa} 
x^{\lambda}\eta^{\mu\nu}x^{2}\\
{}&{} \\
{}&\left.~+k_{6}\eta^{\mu\nu}\delta^{\lambda}_{\kappa}(x^{2})^{2} 
+k_{7}(\eta^{\nu\lambda}\delta^{\mu}_{\kappa}+ 
\eta^{\mu\lambda}\delta^{\nu}_{\kappa})\right\}
\end{array}
\end{equation}
Symmetric part for $\varsigma, \tau$ leads to
\begin{equation}
0=(k_{1}+k_{2}+5k_{3}+k_{4}+k_{5})x^{\nu}x^{\lambda}+(k_{4}+k_{6}+5k
_{7})\eta^{\nu\lambda}x^{2}
\end{equation}
and so 
\begin{equation}
\begin{array}{ll}
k_{1}+k_{2}+5k_{3}+k_{4}+k_{5}=0~~&~
k_{4}+k_{6}+5k_{7}=0
\end{array}
\label{restriction2}
\end{equation}
By  
${\epsilon^{\varsigma\tau}}_{\kappa\mu} 
\epsilon^{\kappa\mu\lambda\nu}=2(\eta^{\varsigma\nu} 
\eta^{\tau\lambda}-\eta^{\varsigma\lambda}\eta^{\tau\nu})$,  
anti-symmetric part leads to
\begin{equation}
\begin{array}{ll}
0=&i\left\{4c_{1}\epsilon^{\varsigma\tau\nu\lambda}x^{2}-
(4c_{1}\epsilon^{\rho\tau\nu\lambda}x^{\varsigma}+ 
4c_{1}\epsilon^{\varsigma\rho\nu\lambda}x^{\tau}+(k_{2}-
k_{4})\epsilon^{\varsigma\tau\rho\lambda}x^{\nu}+(k_{5}-k_{3}) 
\epsilon^{\varsigma\tau\nu\rho}x^{\lambda})x_{\rho}\right\}\\
{}&{}\\
{}&+(4c_{1}-k_{2}+k_{4})
x^{\varsigma}x^{\nu}\eta^{\lambda\tau}+(4c_{1}+k_{3}-
k_{5})x^{\tau} x^{\lambda}\eta^{\varsigma\nu}-(4c_{1}+k_{3}-
k_{5})x^{\varsigma}x^{\lambda}\eta^{\tau\nu}\\
{}&{}\\
{}&-(4c_{1}-k_{2}+k_{4})x^{\tau}x^{\nu}\eta^{\varsigma\lambda}
\end{array}
\label{asp}
\end{equation}
Considering the case when $\varsigma,\tau,\nu,\lambda$ are  all 
different gives
\begin{equation}
\begin{array}{ll}
0=4c_{1}-k_{2}+k_{4}~~~&~~
0=4c_{1}+k_{3}-k_{5}
\end{array}
\label{restriction3}
\end{equation}
and so eq.(\ref{asp}) becomes
\begin{equation}
c_{1}\{(
\epsilon^{\rho\tau\nu\lambda}x^{\varsigma}+ 
\epsilon^{\varsigma\rho\nu\lambda}x^{\tau}+ 
\epsilon^{\varsigma\tau\rho\lambda}x^{\nu}+ 
\epsilon^{\varsigma\tau\nu\rho}x^{\lambda})x_{\rho}-
\epsilon^{\varsigma\tau\nu\lambda}x^{2}\}=0
\end{equation}
which holds automatically regardless of the value of  
$c_{1}$.\newline
\newline
For $\theta^{2}\bar{\theta}$-term.\newline
Direct calculation gives
\begin{equation}
\begin{array}{ll}
0=&(4k_{1}+12k_{3})x^{\nu}x^{\lambda}x\cdot\sigma\\
{}&{}\\
{}&-(k_{1}+k_{2}+3k_{3}+k_{4}+k_{5}+4k_{6}+4k_{7}) 
(x^{\nu}\sigma^{\lambda}+x^{\lambda}\sigma^{\nu})x^{2} \\ 
{}&{}\\
{}&+(k_{2}+k_{3}+3k_{4}+k_{5}+4k_{6}+12k_{7})\eta^{\nu\lambda} 
x\cdot \sigma x^{2}\\
{}&{}\\
{}&+i(4c_{2}-6c_{1}-k_{2}+k_{3}+k_{4}-k_{5}-
4k_{6}+4k_{7})\epsilon^{\nu\lambda\kappa\rho}x_{\kappa} 
\sigma_{\rho}x^{2}
\end{array}
\end{equation}
Various choices of $\nu,\lambda,x^{\nu},x^{\lambda}$ can confirm 
that each term should vanish, hence
\begin{equation}
\begin{array}{l}
0=k_{1}+3k_{3}\\
{}\\
0=k_{2}+k_{4}+k_{5}+4k_{6}+4k_{7}\\
{}\\
0=k_{2}+k_{3}+3k_{4}+k_{5}+4k_{6}+12k_{7}\\
{}\\
0=4c_{2}-6c_{1}-k_{2}+k_{3}+k_{4}-k_{5}-4k_{6}+4k_{7}
\end{array}
\label{restriction4}
\end{equation}
the solution of
eqs.(\ref{restriction1},\ref{restriction2},\ref{restriction3},\ref{restriction4})
may be written in terms of  two parameters,
$c,d$
\begin{equation}
\begin{array}{lllll}
c_{1}=c~~&~c_{2}=\textstyle{\frac{3}{2}}c~~&~
k_{1}=-12d~~&~k_{2}=-6d~~&~k_{3}=4d\\
{}&{}&{}&{}&{}\\
k_{4}=-4c-6d~~&~k_{5}=4c+4d~~&~k_{6}=-c+d~~&~k_{7}=c+d&{}
\end{array}
\end{equation}
which determines the three-point function of conserved supervector 
fields  as eq.(\ref{threesupercurrent}). 
Now we will show that the ``weak''  conservation 
condition~(\ref{3conservation}) actually implies the ``strong''  
one~(\ref{strongconservation}).  By the transformation rule for 
$\partial_{\mu}$~(\ref{partialmu}),
the ``weak'' conservation condition is equivalent to
\begin{equation}
\begin{array}{ll}
0=&{\displaystyle
\left[2(\eta-3)x_{13\kappa}-\Big\{\frac{\partial~}{ \partial 
X_{3}^{\kappa}} 
+(x_{13}^{2}+\theta^{2}_{13}\bar{\theta}^{2}_{13}){I^{-1\mu}}_{\kappa}
(z_{13})\Big(\frac{\partial\Theta_{3}^{\alpha}}{\partial 
x_{13}^{\mu}}\frac{\partial^{-}~}{\partial 
\Theta_{3}^{\alpha}}+\frac{\partial\overline{\Theta}_{3}^{\dot{\alpha}}}
{\partial x_{13}^{\mu}}\frac{\partial^{+}~}{\partial 
\overline{\Theta}_{3}^{\dot{\alpha}}}\Big)\Big\}\right]}\\
{}&{}\\
{}&~{\displaystyle\times\frac{{\cal 
H}^{\kappa\nu\lambda}(Z_{3})}{(X_{3}^{2}+\Theta^{2}_{3} 
\overline{\Theta}^{2}_{3})^{\eta/2}}}
\end{array}
\label{3terms}
\end{equation}
When $\theta_{i}=\bar{\theta}_{i}=0,~i=1,2,3~$,  this equation 
becomes
\begin{equation}
2(3-\eta)x_{13\kappa}\frac{{\cal 
H}^{\kappa\nu\lambda}(Z_{3})}{(X_{3}^{2})^{\eta/3}}=\frac{\partial~}
{\partial X_{3}^{\kappa}}\frac{{\cal 
H}^{\kappa\nu\lambda}(Z_{3})}{(X_{3}^{2})^{\eta/3}}
\end{equation}
Left hand side is a function of $x_{13},X_{3}$ and right hand side 
is a function of $X_{3}$. Hence for this equation to hold for any 
value of $x_{13},X_{3}$,~ $\eta$ should be $3$. Hence 
eq.(\ref{3terms}) becomes
\begin{equation}
{\displaystyle
0=\left\{\frac{\partial~}{ \partial X_{3}^{\kappa}} 
+(x_{13}^{2}+\theta^{2}_{13}\bar{\theta}^{2}_{13}){I
^{-1\mu}}_{\kappa}(z_{13})\Big(\frac{\partial\Theta_{3}^{\alpha}}{
\partial x_{13}^{\mu}}\frac{\partial^{-}~}{\partial 
\Theta_{3}^{\alpha}}+\frac{\partial\overline{\Theta}_{3}^{\dot{ 
\alpha}}}{\partial x_{13}^{\mu}}\frac{\partial^{+}~}{\partial 
\overline{\Theta}_{3}^{\dot{\alpha}}}\Big)\right\}\frac{{\cal 
H}^{\kappa\nu\lambda}(Z_{3})}{(X_{3}^{2}+\Theta^{2}_{3} 
\overline{\Theta}^{2}_{3})^{3/2}}}
\label{3degenerate}
\end{equation}
$(x_{13}^{2}+\theta^{2}_{13}\bar{\theta}^{2}_{13}){I^{-
1\mu}}_{\kappa}
(z_{13})\frac{\partial\Theta_{3}^{\alpha}}{\partial x_{13}^{\mu}}$
is a function of 
$x_{13},\bar{\theta}_{13},\theta_{13}\bar{\theta}^{2}$ and 
$(x_{13}^{2}+\theta^{2}_{13}\bar{\theta}^{2}_{13}){I^{-
1\mu}}_{\kappa}
(z_{13})\frac{\partial\overline{\Theta}_{3}^{\dot{\alpha}}}{\partial 
x_{13}^{\mu}}$ is a function of 
$x_{13},\theta_{13},\bar{\theta}_{13}\theta^{2}_{13}$. Thus each of 
the three terms appearing  in eq.(\ref{3degenerate}) should be zero and  
so 
\begin{eqnarray}
&{\displaystyle 0=\partial_{\kappa}\frac{{\cal 
H}^{\kappa\nu\lambda}(z)}{(x^{2}+\theta^{2}\bar{\theta}^{2})^{3/2}}}
~~~\label{first}\\
\nonumber\\
&{\displaystyle0=\tilde{\sigma}^{\dot{\alpha}\alpha}_{\kappa} 
\partial^{-}_{\alpha}\frac{{\cal 
H}^{\kappa\nu\lambda}(z)}{(x^{2}+\theta^{2}\bar{\theta}^{2})^{3/2}}}
\label{second}\\
\nonumber\\
&{\displaystyle0=\tilde{\sigma}^{\dot{\alpha}\alpha}_{\kappa} 
\bar{\partial}^{+}_{\dot{\alpha}}\frac{{\cal 
H}^{\kappa\nu\lambda}(z)}{(x^{2}+\theta^{2}\bar{\theta}^{2})^{3/2}}}
\label{third}
\end{eqnarray}
Nevertheless eq.(\ref{third}) is just the ``strong'' conservation 
equation. The other two equations can be obtained from 
eq.(\ref{third}) by taking its complex conjugate and using 
$\{\partial^{-}_{\alpha},\bar{\partial}^{+}_{\dot{\alpha}}\}=2i
\partial_{\alpha\dot{\alpha}}$.

\bibliographystyle{unsrt}
\bibliography{reference}

\end{document}